\documentclass[10pt]{article}

\usepackage[left=2.5cm,right=2.5cm,top=2.8cm,bottom=2.8cm]{geometry}

\usepackage{hyperref}
\usepackage{amsmath}
\usepackage{amssymb}
\usepackage{amsthm}
\usepackage{listings}
\usepackage[ruled,vlined,linesnumbered]{algorithm2e}
\usepackage{bm}
\usepackage[skip=0cm,list=true,labelfont=it]{subcaption}
\usepackage{mathtools}
\usepackage{multirow}
\usepackage{booktabs}
\usepackage{xcolor}

\newcommand{\nop}[1]{}
\newcommand{\lift}{\lambda}
\newcommand{\liftcon}{\lift\textsubscript{con}}
\newcommand{\liftcat}{\lift\textsubscript{cat}}
\renewcommand{\vec}[1]{\textbf{#1}}
\newcommand{\punto}{$\hspace*{\fill}\Box$}
\newcommand{\bigO}[1]{\mathcal{O}(#1)}

\newcommand\hl[1]{#1}
\DeclareMathOperator{\diag}{diag}
\DeclareMathOperator*{\argmax}{argmax}

\theoremstyle{definition}
\newtheorem{example}{Example}

\providecommand{\keywords}[1]{\textbf{{Keywords:}} #1}

\lstdefinelanguage{SQL}{
  basicstyle=\linespread{1.05}\ttfamily,
  morekeywords={SELECT, FROM, NATURAL, JOIN, GROUP, BY, ON, AS, WHERE},
  keywordstyle=\linespread{1.05}\ttfamily\color{blue!45!black},
  tabsize=2,
  numbers=none,
  stringstyle=\color{white}\ttfamily,
  showspaces=false,
  showtabs=false,
  showstringspaces=false
}

\title{In-Database Data Imputation}

\author{Massimo Perini \and Milos Nikolic}

\date{
	\{massimo.perini, milos.nikolic\}@ed.ac.uk \\[12pt]
	University of Edinburgh
}

\begin{document}

\maketitle

\begin{abstract}
	Missing data is a widespread problem in many domains, creating challenges in data analysis and decision making. Traditional techniques for dealing with missing data, such as excluding incomplete records or imputing simple estimates (e.g., mean), are computationally efficient but may introduce bias and disrupt variable relationships, leading to inaccurate analyses. Model-based imputation techniques offer a more robust solution that preserves the variability and relationships in the data, but they demand significantly more computation time, limiting their applicability to small datasets.
	
	This work enables efficient, high-quality, and scalable data imputation within a database system using the widely used MICE method. We adapt this method to exploit computation sharing and a ring abstraction for faster model training. To impute both continuous and categorical values, we develop techniques for in-database learning of stochastic linear regression and Gaussian discriminant analysis models. Our MICE implementations in PostgreSQL and DuckDB outperform alternative MICE implementations and model-based imputation techniques by up to two orders of magnitude in terms of computation time, while maintaining high imputation quality.
\end{abstract}


\section{Introduction}\label{sec:introduction}

Missing data is pervasive in real-world datasets across various domains, posing significant challenges in drawing accurate and reliable conclusions from incomplete information.
Missing data may introduce bias, reduce statistical power, and undermine the validity of analyses and predictive models~\cite{Stoyanovich:2020:Responsible, Little:2019:Book, Graham:2009:RevPsyc}.
Moreover, traditional machine learning techniques, typically designed to operate on complete datasets, are ill-equipped to handle missing data effectively~\cite{Emmanuel:2021:Survey,  Jager:2021:Frontiers, Kumar:2017:Tutorial, Toussaint:2022:VLDB}.
Therefore, handling missing data effectively is essential for the validity and robustness of data analysis.


\begin{figure}[t]
	\centering
	\includegraphics[width=0.55\linewidth]{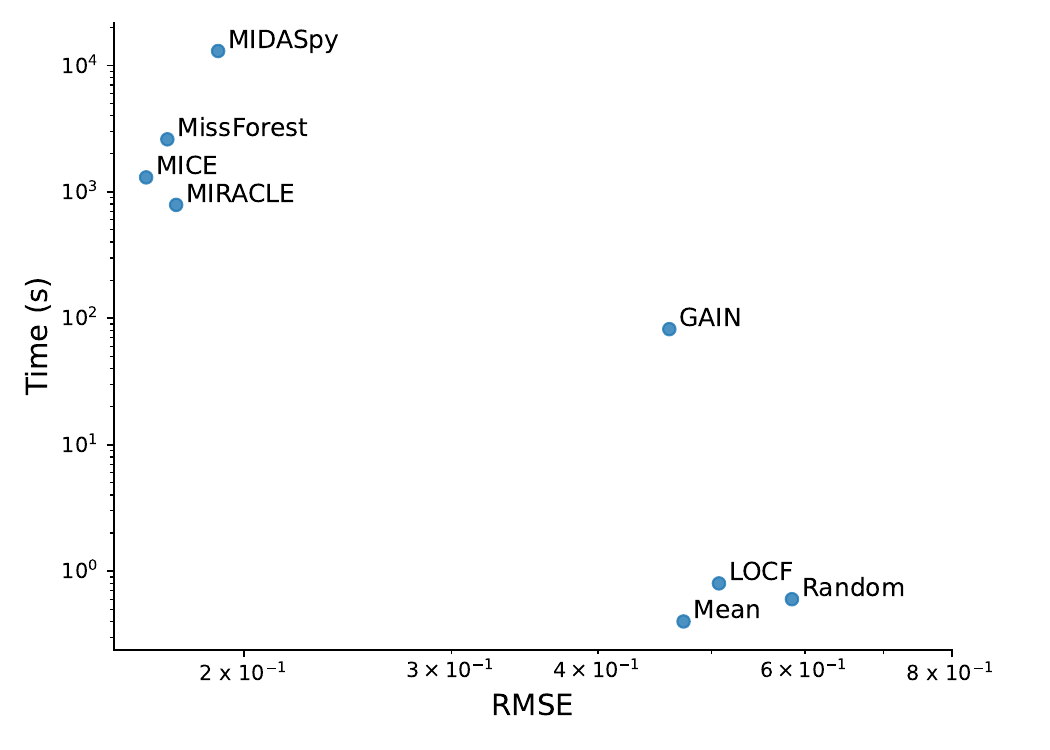}
	\caption{
		\hl{Imputation quality and runtime of Python-based imputation methods on a flight dataset~\cite{dataset:flight} with 5M rows and 20\% missing values. Imputation quality is measured as the root mean square error (RMSE) of the linear regression model trained over an imputed dataset to predict flight duration.}
	}
	\label{fig:comp_imputation}
\end{figure}

Various techniques are commonly used to handle datasets with missing information, including disregarding records with missing data, incorporating indicator variables to capture the missing pattern, 
and imputing alternative values in place of the missing values. 
Common imputation methods involve mean imputation, which replaces missing values with the mean of the column; last observation carried forward (LOCF), which substitutes missing values with the most recent observed value~\cite{Lachin:2016:LOCF}; and hot deck imputation, which utilizes similarity criteria within the column~\cite{Andridge:2010:HotDeck}. Although these techniques are easy to implement and computationally efficient, they offer limited guarantees regarding the quality of the imputed data. They can potentially distort value distributions, underestimate data variance, disrupt variable relationships, and introduce bias in statistical measures such as the mean, leading to subpar analytics and machine learning models over the imputed data~\cite{Stoyanovich:2020:Responsible,Chu:2016:DataCleaning, vanBuuren:2018:Book, Lachin:2016:LOCF}.

Model-based imputation methods can overcome these problems and capture complex missingness patterns while preserving the underlying data distribution.
\hl{
These methods learn models over observed data to impute missing data.
Popular iterative imputation methods include MissForest~\cite{Stekhoven:2012:MissForest}, which utilizes random forests, and
Multivariate Imputation by Chained Equations (MICE)~\cite{vanBuuren:2011:miceR}, which allows using a different model for each attribute with missing values.
Recent generative imputation methods exploit generative adversarial networks, like in GAIN~\cite{Yoon:2018:GAIN}, and deep learning, like in MIDASpy~\cite{Ranjit:2022:MIDAS} and MIRACLE~\cite{Kyono:2021:Miracle}.
Despite their effectiveness, such model-based methods suffer from high computation costs.
Figure~\ref{fig:comp_imputation} illustrates the trade-off between the quality of imputation, measured by the performance of a linear regression model trained on imputed data, and the time needed for imputation using different methods on a real dataset.
Compared to model-free imputation methods such as mean and random imputation, 
complex methods such as MICE and MIRACLE yield models with lower root mean squared errors (RMSE), indicating higher quality, but demand orders of magnitude more time for model training and data imputation.
}

We next identify the challenges involved in enabling high-quality imputation on large datasets. 

{\bf Challenge 1: Enabling Model-Based Imputation in DBMSs.}
Model-based imputation methods are commonly implemented in external tools such as scikit-learn~\cite{scikit-learn} and R~\cite{vanBuuren:2011:miceR}, operating independently of the DBMS environment. 
To impute a dataset stored in a DBMS using these methods, we need to export the dataset to the external tool for imputation and import the imputed dataset back into the DBMS for further analyses.
These steps can significantly increase the computation time. 
Moreover, as the external tool typically lacks support for out-of-memory computation, this approach works only with datasets small enough to fit into memory.

{\bf Challenge 2: Long Imputation Time.}
Despite the benefits offered by model-based imputation methods, simpler techniques such as dropping rows with missing values and mean imputation continue to be prevalent in practice and are often the default option in statistical packages such as R and SAS. The reluctance to adopt more complex imputation methods primarily stems from their higher computational time and limited ability to handle large datasets~\cite{Graham:2009:RevPsyc,vanBuuren:2018:Book,Shafer:2002:PhysMet}.
For instance, MissForest and MICE impute missing data by employing an iterative process of predictive modeling~\cite{Graham:2009:RevPsyc,Stekhoven:2012:MissForest}, retraining models from scratch on every iteration. Data practitioners utilizing complex imputation methods may experience extended waiting periods, potentially lasting several hours, for the imputation process to converge~\cite{Cambronero:2017}. Therefore, these complex methods are typically viable only for small datasets.

{\bf Challenge 3: Avoiding Data Explosion.}
Model-based imputation methods implemented in external tools can only train models on data available in a single table. Therefore, data practitioners need to preprocess the dataset before exporting it to the external tool, which may involve joining relations if the dataset is normalized and performing one-hot encoding of categorical features.
These preprocessing steps introduce redundancy in both data and computation, significantly increasing the size of the training dataset and prolonging the overall computation time~\cite{Schleich:2016:Learning, Chen:2017:Morpheus, Schleich:2019:LMFAO}.

\hl{
\textbf{Our Goal.} 
This work aims to enable efficient and high-quality data imputation within database systems. By moving imputation closer to data, we want to harness the performance and scalability of database systems to accelerate both model training and data imputation, effectively addressing the above challenges.
}

\hl{
This work studies in-database data imputation using the MICE method~\cite{vanBuuren:2018:Book,vanBuuren:2011:miceR}.
This method can impute incomplete multivariate data comprising mixtures of continuous and categorical values.
For each attribute with missing values, MICE learns a model -- a regressor or a classifier depending on the attribute's type -- using all other attributes as predictors.
The imputation proceeds one attribute at a time in a round-robin fashion, incorporating previously imputed values into the prediction models for subsequent attributes.
}

\hl{
We focus on MICE for three compelling reasons:
1) MICE offers flexibility and can utilize various predictive models, including those that can be efficiently trained inside database systems;
2) In terms of imputation quality, MICE competes with or outperforms other state-of-the-art imputation methods, including recent generative approaches~\cite{Zheng:2022:Diffusion,Yoon:2018:GAIN}; 
3) MICE has been extensively studied in the past, widely used across diverse domains~\cite{vanBuuren:2011:miceR, Mera-Gaona:2021:Feature, Huque:2018:Comparison, Ambler:2007:RiskModeling}, and implemented in popular statistical tools~\cite{vanBuuren:2011:miceR, Yuan:2010:SAS}.
We leave studying other model-based imputation methods as future work.
}

\textbf{Our contributions.} 
\hl{We next highlight our main contributions. }




To support MICE within a database system (Challenge 1), 
\hl{
we introduce techniques for in-database training of two novel models:
stochastic linear regression~\cite{vanBuuren:2018:Book}, used for imputing continuous values, and
Gaussian discriminant analysis, used for imputing categorical values (Section~\ref{sec:learning}). 
}
The former builds upon existing work on learning regression models~\cite{Schleich:2016:Learning, Nikolic:2018:SIGMOD} by incorporating random noise into predictions to capture the uncertainty associated with the imputation process.  
\hl{
The latter transforms a multinomial classification problem into a database problem, computing the model parameters using a single database query over normalized data, without the need for prior one-hot encoding and materialization of the training dataset. 
To the best of our knowledge, no prior work has explored classification methods in this setting.
}
Interestingly, both models compute the same set of database aggregates in the training phase, despite their distinct characteristics.


To reduce imputation time (Challenge 2), we adapt the MICE algorithm to exploit computation sharing across iterations while preserving its accuracy, \hl{regardless of the proportion of missing data} (Section~\ref{sec:indb-imputation}).
To unlock the sharing potential, we select stochastic linear regression and Gaussian discriminant analysis as the models for imputing continuous and categorical attributes, respectively. 
As mentioned, training these models within a database system relies on computing the same set of aggregate values. 
We observe that MICE computes these aggregates over overlapping subsets of records.
\hl{
The extent of overlap increases as the proportion of missing values decreases. For datasets with low missing rates, we propose a rewriting of the MICE algorithm that:
}
1) performs the most expensive computation over the entire dataset once, outside the iterative loop; and 
2) performs less expensive incremental computations over (smaller) incomplete parts of the dataset inside the iterative loop.
\hl{
This rewriting accelerates model training over datasets with less than 20\% missing values, based on our experimental results. 
We further devise partitioning strategies to
minimize redundant computation and
avoid repetitive scans of the entire dataset, tailoring to both low and high missing rate scenarios.
}
Our improvements of the MICE algorithm are applicable not only within a database system but also in other tools and libraries implementing MICE.


\nop{
We address the second challenge by rewriting the MICE algorithm such that we can exploit computation sharing. As can be seen from the example in Fig. \ref{fig:mice}, in each iteration most of the rows are used multiple times in different models. Assuming these rows are the majority, we notice how having models trained on similar sets of rows leads to training sets which overlap with each other, causing redundant computation. 

We therefore rewrite the MICE algorithm such that calculations over fixed data are performed once and shared across models and iterations. Decoupling the parameter convergence phase from the computation of aggregate values allows to perform expensive computations over the dataset once and efficiently update these aggregates when imputed values change.
}

To avoid materialization of large datasets (Challenge 3),
we optimize the computation of MICE aggregates 
\hl{using the mathematical notion of ring\footnote{\hl{A ring is a set $\mathcal{R}$ with two binary operations, $+$ (addition) and $*$ (multiplication), such that $(\mathcal{R}, +)$ is an abelian group, $(\mathcal{R}, *)$ is a monoid, and $*$ is distributive over $+$.}}}.
We leverage the cofactor ring from prior work~\cite{Nikolic:2018:SIGMOD,Nikolic:2020:FIVM} to compactly encode the needed aggregates as ring values and compute them using the sum and product from the ring. 
When imputing a normalized dataset, we exploit the algebraic properties of the ring to push the aggregate computation past joins, eliminating the need to materialize the joined relation. 
Furthermore, to avoid the size explosion caused by one-hot encoding of categorical attributes, 
we utilize the generalized cofactor ring~\cite{Nikolic:2020:FIVM} to uniformly encode the needed aggregates over continuous and categorical attributes as generalized multiset relations~\cite{Koch:2010:PODS}. This representation accounts for only the interactions between attributes that exist in the dataset, avoiding the sparsity of one-hot encoding. 
Finally, the ring-based representations is essential for our optimized MICE implementation because it allows the aggregates to be incrementally computed as new values are imputed.

We implemented our data imputation approach in PostgreSQL and DuckDB, including the procedures for in-database learning of  ridge linear regression, stochastic linear regression, and Gaussian discriminant analysis models.
To efficiently compute the aggregates required for training these models, our implementation harnesses the power of the cofactor ring,
representing the first instance of incorporating this abstraction into fully-fledged database systems.
Our ring-based implementation improves the performance of computing these aggregates in PostgreSQL and DuckDB by up to 6x over a single table and up to 12x over multiple tables. 

Our experiments show that our in-database imputation outperforms the fastest competitor,
SystemDS~\cite{Boehm:2016:SystemML, system:SystemDS}, by 3-13x when using PostgreSQL and 86-346x when using DuckDB, depending on the fraction of missing values. 
For datasets comprising multiple tables, our imputation method can leverage factorized evaluation to bring a 6x improvement compared to imputing over the joined table. 
Our MICE method is on par with other model-based methods in terms of imputation quality but 
\hl{
requires up to two orders of magnitude less time, under various missing rates and patterns.
}

The rest of the paper is organized as follows. 
Section~\ref{sec:background} introduces background material.
Section~\ref{sec:learning} shows how to support stochastic linear regression and Gaussian discriminant analysis within a database system. 
Section~\ref{sec:indb-imputation} presents our adapted MICE algorithm and its implementation in Section~\ref{sec:implementation}.
Section~\ref{sec:experiments} gives experimental results, followed by a discussion of related work and conclusion. 

\hl{Our implementation of the above machine learning and imputation methods in PostgreSQL and DuckDB is publicly available at \url{https://github.com/eddbase/db-imputation}.}

\section{Background}\label{sec:background}

We next introduce the MICE algorithm for imputing missing values and review the relevant prior work on in-database learning.

\hl{
{\bf Notation.\ }
Consider an incomplete dataset $\vec{X}$ consisting of records over attributes $X_1, \ldots, X_m$. 
We denote by $\tilde{\vec{X}}$ a complete version of $\vec{X}$ with the missing values replaced by imputed values.
We write $\tilde{\vec{X}}_{i = \mathit{miss}}$ (or $\tilde{\vec{X}}_{i = \mathit{obs}}$) to denote the records from $\tilde{\vec{X}}$ for which the value of attribute $X_i$ was originally missing (or observed).
}

{
\begin{algorithm}[t]	
	\caption{MICE} \label{alg:mice_v2}
	\SetAlgoLined
	\SetKwInput{Input}{Input\hspace{0.7em}}
    \SetKwInput{Output}{Output}
	\SetKwProg{Fn}{Function}{:}{}	
	\Input{
		\begin{tabular}[t]{ll@{}}
			$\vec{X}$ & incomplete dataset with attributes $X_1, ..., X_m$ \\ 
			$\mathit{mattrs}$ & indices of incomplete attributes
		\end{tabular} 
	}	
	\Output{
		\begin{tabular}[t]{ll}
			$\tilde{\vec{X}}$ \hspace{1.6em} & imputed dataset
		\end{tabular}
	}
	{
		$\tilde{\vec{X}} \leftarrow \vec{X}$ with initial imputations for all missing values\\
		\Repeat{ $\mathit{stopping\_condition}$}{
			\ForEach{$i \in \mathit{mattrs}$}{
				$\bm{\theta} \leftarrow$ \textsc{Train}($\mathit{data} = \tilde{\vec{X}}_{i=\mathit{obs}}$,\, $\mathit{target} = X_i$) \\
				$\tilde{\vec{X}} \leftarrow \textsc{Predict}(\mathit{data} = \tilde{\vec{X}}_{i=\mathit{miss}},\, \mathit{target} = X_i, \mathit{model} = \bm{\theta})$ \\
			}
		}
	}
\end{algorithm}
}

\subsection{The MICE Algorithm}

Multivariate Imputation by Chained Equations (MICE) is an imputation method for handling missing data in multivariate datasets~\cite{vanBuuren:2011:miceR}. 
The method iteratively imputes missing values in each attribute based on the observed values in the other attributes.
\hl{This imputation process accounts for the patterns present in the data, preserving correlations among attributes and yielding  more accurate imputations than model-free methods (e.g., mean imputation).}

The MICE method, shown in Algorithm~\ref{alg:mice_v2}, starts by imputing the missing data with initial guesses, typically the mean/mode values. 
For each attribute $X_i$ with missing values, 
the algorithm fits a model over the observed part $\tilde{\vec{X}}_{i=\mathit{obs}}$ with $X_i$ as target and then uses this model to generate new imputations of $X_i$ in the missing part $\tilde{\vec{X}}_{i=\mathit{miss}}$.
This iterative process continues until a stopping criterion is satisfied or the maximum number of iterations specified by the user is reached. 
The MICE algorithm offers flexibility by allowing the use of distinct models for different attributes, making it capable of imputing missing values of both continuous and categorical attributes.

We refer to one round of imputations of all incomplete attributes as one iteration step. While the MICE algorithm offer no  convergence guarantees, it generally converges within a small number of iterations in practice, typically between 5 and 20 iterations~\cite{vanBuuren:2018:Book}.

\begin{example}\label{ex:mice}
\hl{
	Consider an incomplete flight dataset with three attributes, two continuous ($\mathit{Distance}$ and $\mathit{Air Time}$) and one categorical ($\mathit{Diverted}$), each with missing values. 
	The MICE algorithm first imputes the missing values in each attribute with the mean or mode, depending on the attribute's type.
	It then trains a regression model to predict $\mathit{Distance}$ given $\mathit{Air Time}$ and $\mathit{Diverted}$ over the subset of records with complete $\mathit{Distance}$ values. The trained model serves to impute missing $\mathit{Distance}$ values.  
	The second regression model predicts $\mathit{Air Time}$, while the third classification model predicts $\mathit{ Diverted}$.
	The algorithm refines the imputed values in a round-robin way until convergence or a fixed number of iterations. 
}
	\punto
\end{example}

\nop{
\begin{example}
	We demonstrate one iteration step of the MICE algorithm on a toy dataset containing flight information. 
	Figure~\ref{fig:mice} shows the initial dataset with three incomplete attributes: two continuous ($\mathit{Distance}$ and $\mathit{Air Time}$) and one categorical ($\mathit{Diverted}$).
	
	The algorithm first replaces the missing values in each attribute with the mean or mode, depending on the attribute's type.
	It then iteratively improves these estimates using predictive models with one attribute as target and the other attributes as predictors. 
	The algorithm trains two regression models for predicting $\mathit{Distance}$ and $\mathit{Air Time}$ and one classifier for predicting $\mathit{Diverted}$.
	The training datasets are marked in yellow, while the imputed values are marked in orange.
	The algorithm refines the imputed values in a round-robin way until convergence or a fixed number of iterations. 
	\punto
\end{example}
}

The MICE algorithm involves retraining models for every incomplete attribute in every iteration, making it potentially prohibitively expensive on large datasets with many incomplete attributes. 

\subsection{In-Database Linear Regression}\label{sec:linear_regression}

Consider a training dataset $\vec{X}$ consisting of $N$ training examples with attributes $X_1, \ldots, X_m$. Without loss of generality, assume that $X_1 = 1$ for all examples in $\vec{X}$, and $X_m$ is the target attribute.
The goal of linear regression is to find the parameters $\bm{\theta} = [ \theta_1 \ldots \theta_{m}]^T$, with $\theta_m=-1$, that minimize the squared error loss ${L} = (\vec{X}\bm{\theta})^T\vec{X}\bm{\theta}$.

Batch gradient descent solves this optimization problem by iteratively updating the learned parameters in the opposite direction of the gradient of $L$ until convergence, while $\theta_m$ is fixed to $-1$:
\begin{equation*}
	\bm{\theta}^{(k+1)} = \bm{\theta}^{(k)} - \alpha \vec{X}^T (\vec{X}\bm{\theta}^{(k)})
	\label{eq:BGD_simple}
\end{equation*}
where $\alpha$ is the learning rate and $\bm{\theta}^{(k)}$ are the parameter values in iteration $k$. 
Repeatedly scanning the complete dataset to compute $\vec{X}^T (\vec{X}\bm{\theta}^{(k)})$ can be time-consuming, especially with large datasets.

A more efficient approach precomputes $\vec{X}^T \vec{X}$ once and reuses it in each iteration, effectively decoupling the computation over the training dataset from the parameter convergence~\cite{Schleich:2016:Learning}:
\hl{
\begin{equation*}
\bm{\theta}^{(k+1)} = \bm{\theta}^{(k)} - \alpha \;\vec{C}\; \bm{\theta}^{(k)}  
\end{equation*}
where $\vec{C} = \vec{X}^T \vec{X}$ is the cofactor matrix of size $m \times m$, which quantifies the level of correlation for each combination of attributes. The one-off computation of $\vec{X}^T \vec{X}$ takes $\bigO{Nm^2}$ time.
Using the precomputed cofactor matrix, 
each iteration now takes time $\bigO{m^2}$ instead of $\bigO{Nm}$, yielding faster convergence as usually $N \gg m$.
}

{\bf Cofactor Matrix Computation.}
Let us first assume that all attributes are continuous. 
The cofactor matrix $\vec{X}^T \vec{X}$ accounts for the interactions \texttt{SUM(X\textsubscript{i}\,*\,X\textsubscript{j})} of all pairs $(X_i, X_j)$ of attributes. 
Thus, computing the cofactor matrix amounts to executing a database query with $\bigO{m^2}$ sum aggregates over the training dataset.

When the training dataset $\vec{X}$ is the result of a join, a na\"{i}ve way of computing the cofactor matrix is to first calculate the join result and then calculate the cofactor aggregates. Based on our experimental evaluation, existing query optimizers make no attempts to factorize the evaluation of many aggregates, that is, to perform partial preaggregation by pushing \texttt{SUM}s past joins~\cite{Larson:2002:ICDE}.  

Prior work shows how to express the cofactor matrix computation as the computation of one compound aggregate. 
This aggregate is a triple $(N, \vec{s}, \vec{Q})$, where 
$N$ is the size of the dataset, \texttt{SUM(1)};
$\vec{s}$ is a vector of sums of values for each attribute, $\vec{s}_i = \texttt{SUM(X\textsubscript{i})}$; and 
$\vec{Q}$ is a matrix of sums of products of values for any two attributes, $\vec{Q}_{(i,j)} = \texttt{SUM(X\textsubscript{i}\,*\,X\textsubscript{j})}$. 
The computation over triples is captured by the cofactor (degree-$m$ matrix) ring~\cite{Nikolic:2020:FIVM,Nikolic:2018:SIGMOD}. 

The cofactor ring defines the addition and multiplication operations over triples. 
Let $\mathcal{R}$ be a set of triples $(\mathbb{Z}, \mathbb{R}^m,\mathbb{R}^{m \times m})$, for fixed $m\in\mathbb{N}$. For any $a = (N_a, \vec{s}_a, \vec{Q}_a) \in \mathcal{R}$ and $b = (N_b, \vec{s}_b, \vec{Q}_b) \in \mathcal{R}$, the addition and multiplication operations on $\mathcal{R}$ are defined as:
\begin{align*}
	a +^\mathcal{R} b &= (N_a + N_b,\ \vec{s}_a + \vec{s}_b,\ \vec{Q}_a + \vec{Q}_b) \\
	a *^\mathcal{R} b &= (N_a\, N_b,\ N_b\,\vec{s}_a + N_a \,\vec{s}_b,\ N_b \,\vec{Q}_a + N_a\,\vec{Q}_b + \vec{s}_a \vec{s}_b^T + \vec{s}_b \vec{s}_a^T)
\end{align*} 
where $+^\mathcal{R}$ uses scalar and matrix addition,
and $*^\mathcal{R}$ uses matrix addition and scalar and matrix multiplication. 
The additive identity (zero) is 
$(0, \vec{0}_{m\times 1}, \vec{0}_{m\times m})$ and 
the multiplicative identity (one) is 
$(1, \vec{0}_{m\times 1}, \vec{0}_{m\times m})$,
where $\vec{0}_{m\times n}$ is the zero matrix of size $m \times n$.

The function $\lift_{\text{con}}$ maps values of continuous attribute $X$ to triples from $\mathcal{R}$ such that
$\lift_{\text{con}}(x,i) = (1, \vec{s}, \vec{Q})$, 
where $i$ is the index of $X$ in the cofactor matrix, and
 $\vec{s}$ and $\vec{Q}$ contain all zeros except $\vec{s}_i = x$ and $\vec{Q}_{(i,i)} = x^2$. 
 We refer to $\lift_{\text{con}}$ as a lifting function. 
 For brevity, we omit the index $i$ in $\lift_{\text{con}}$, assuming a fixed order of attributes.

\begin{example}
	For the flight dataset from Example~\ref{ex:mice}, 
	we can compute the cofactor matrix over the attributes $\mathit{Distance}$ and $\mathit{Air Time}$ using one aggregate query, which returns a triple from the cofactor ring:

\begin{lstlisting}[language=SQL, mathescape, columns=fullflexible]
    SELECT SUM($\lift\textsubscript{con}\hspace{-0.1em}$(Distance)$\,$*$\,\lift\textsubscript{con}\hspace{-0.1em}$(AirTime)) FROM Flight
\end{lstlisting}
The function $\lift_{\text{con}}$ maps a $\mathit{Distance}$ value $d$ to $(1, \begin{bsmallmatrix}d & 0\end{bsmallmatrix}, \begin{bsmallmatrix}d^2 & 0 & ; & 0 & 0\end{bsmallmatrix})$ and similarly an $\mathit{AirTime}$ value $a$ to $(1, \begin{bsmallmatrix}0 & a\end{bsmallmatrix}, \begin{bsmallmatrix}0 & 0 & ; & 0 & a^2\end{bsmallmatrix})$. 
The triple multiplication yields $(1, \begin{bsmallmatrix}d & a\end{bsmallmatrix}, \begin{bsmallmatrix}d^2 & da & ; & ad & a^2\end{bsmallmatrix})$. 
The \texttt{SUM} operator uses the addition $+^\mathcal{R}$ from the cofactor ring. 
The result encodes the cofactor aggregates: 
\texttt{SUM(1)}, 
\texttt{SUM(D)}, \texttt{SUM(A)},
\texttt{SUM(D*D)}, \texttt{SUM(D*A)}, and \texttt{SUM(A*A)},
where \texttt{D} and \text{A} stand for \texttt{Distance} and \texttt{AirTime}, respectively.
	\punto
\end{example}

When computing the cofactor matrix over joins, 
this ring-based approach allows for the \texttt{SUM} to be pushed past the joins, leveraging the distributivity of multiplication over addition in $\mathcal{R}$. 
This eliminates the need to compute the entire join result in advance and promotes the sharing of aggregate computation.

{\bf Handling Categorical Attributes.}
Most machine learning algorithms require numerical input and cannot work directly with discrete categories. 
One-hot encoding is often employed to represent categorical attributes as indicator vectors.
But this encoding step can cause data explosion as each category yields a new binary attribute.

The cofactor matrix of a one-hot encoded dataset comprises the following aggregates:
\texttt{SUM(X\textsubscript{i}\,*\,X\textsubscript{j})} when $X_i$ and $X_j$ are continuous;
\texttt{SUM(X\textsubscript{i})$\;$group$\;$by$\;$X\textsubscript{j}} when $X_i$ is continuous and $X_j$ is categorical;
and 
\texttt{SUM(1)$\;$group$\;$by$\;$X\textsubscript{i},$\,$X\textsubscript{j}} when $X_i$ and $X_j$ are categorical~\cite{Schleich:2016:Learning}.

To avoid one-hot encoding and operate directly over categorical values, prior work~\cite{Nikolic:2020:FIVM} generalizes the cofactor ring 
with the ring over relations~\cite{Koch:2010:PODS}
to uniformly treat aggregates with continuous and categorical attributes. 
The triple structure remains unchanged except that $N$, $\vec{s}$, and $\vec{Q}$ contain relations instead of scalars. Each relation is a mapping from tuples to scalars. 
The operations $+^\mathcal{R}$ and $*^\mathcal{R}$ remain unchanged except that scalar addition is replaced by union and scalar multiplication is replaced by join. 
A scalar $c$ is represented as the relation $\{ () \mapsto c \}$ mapping the empty tuple to $c$. 

The generalized cofactor ring requires new functions for mapping attribute values to ring values. 
For a categorical attribute $X$ with index $i$ in the cofactor matrix,  
the lifting function $\lift_{\text{cat}}$ maps categories of $X$ to triples such that
$\lift_{\text{cat}}(x) = (\bm{1}, \vec{s}, \vec{Q})$, 
where $\bm{1}$ is $\{ () \mapsto 1 \}$,
and $\vec{s}$ and $\vec{Q}$ contain all empty relations except $\vec{s}_i = \{ x \mapsto 1 \}$ and $\vec{Q}_{(i,i)} = \{ x \mapsto 1 \}$. The lifting function $\lift_{\text{con}}$ returns a triple as before but with every scalar $c$ replaced by $\{ () \mapsto c \}$.

\begin{example}
	For the flight dataset from Example~\ref{ex:mice}, 
	we can compute the cofactor matrix over $\mathit{Air Time}$ (continuous) and $\mathit{Diverted}$ (categorical) using a query over the generalized cofactor ring:
\begin{lstlisting}[language=SQL, mathescape, columns=fullflexible]
    SELECT SUM($\lift\textsubscript{con}\hspace{-0.1em}$(AirTime)$\,$*$\,\lift\textsubscript{can}\hspace{-0.1em}$(Diverted)) FROM Flight
\end{lstlisting}
Each tuple $(a,d)$ over the two attributes is mapped to $(N, \vec{s}, \vec{Q})$,
where
$N =\{() \mapsto 1\}$,
$\vec{s} = 
\begin{bsmallmatrix}
\{ () \mapsto a \} & \{ d \mapsto 1 \}
\end{bsmallmatrix}$, and
$\vec{Q} =  
\begin{bsmallmatrix}
	\{ () \mapsto a^2 \} & \{ d \mapsto a \} & ; & 
	\{ d \mapsto a \} & \{ d \mapsto 1 \}
\end{bsmallmatrix}	
)$.
The final aggregate $(N, \vec{s}, \vec{Q})$ encodes:
$\texttt{SUM(1)}$ in $N$;
$\texttt{SUM(A)}$ and
$\texttt{SUM(1)\;group\;by\;D}$ in $\vec{s}$;
$\texttt{SUM(A\,*\,A)}$ and
$\texttt{SUM(A)\;group\;by\;D}$ in $\vec{Q}$,
where \texttt{A} and \text{D} stand for \texttt{AirTime} and \texttt{Diverted}, respectively.
\punto
\end{example}

\section{In-Database Imputation Methods}\label{sec:learning}

This section presents two methods for learning models needed for the imputation of continuous and categorical values, within a database management system. 
The first method extends the approach for learning regression models from Section~\ref{sec:linear_regression} to incorporate random noise into predictions. 
The second method is a novel approach for in-database classification using Gaussian discriminant analysis. 
Seemingly disparate, both methods rely on the aggregates that can be computed using the generalized cofactor ring. 


\nop{
We consider regression and classification tasks that are amenable to in-database computation using ring abstraction:

\begin{itemize}
    \item Ridge Linear Regression, where we extend prior work~\cite{Schleich:2016:Learning,Nikolic:2018} with a generalized ring for computing aggregates needed for training over both continuous and categorical attributes and we introduce Stochastic Linear Regression models.
    \item Generative classification, where we leverage the generalized ring for training Gaussian Discriminant Analysis (GDA) models.
\end{itemize}

Seemingly disparate, both tasks rely on the same set of aggregates during the training phase.
}

\subsection{Stochastic Linear Regression}\label{sec:learning_regression}

Using linear regression for data imputation carries the risk of overstating the strength of the relationship between the target and predictor attributes.
Stochastic regression imputation~\cite{vanBuuren:2018:Book} is a refinement of regression imputation that attempts to address this correlation bias by adding noise to the predictions.
By doing so, imputed data will randomly deviate from the regression line, capturing the inherent uncertainty of the imputation process.

For convenience, we summarize the setup from Section~\ref{sec:linear_regression}. A training dataset $\vec{X}$ consists of $N$ training examples over continuous attributes $X_1, \ldots, X_m$, where $X_m$ denotes the target and $X_1$ is fixed to $1$. 
Linear regression learns the parameters $\bm{\theta} = [ \theta_1, \ldots, \theta_m ]^T$, with $\theta_m = -1$ , minimizing the squared error loss ${L} = (\vec{X}\bm{\theta})^T\vec{X}\bm{\theta}$. 

Stochastic linear regression estimates the parameters $\bm{\theta}$ under the linear model but adds random noise to predictions:
$f(\vec{x}) = \vec{x}^T\bm{\theta'} + \epsilon$, where $\vec{x} \in \mathbb{R}^{m-1}$, $\bm{\theta}' = [\theta_1 \ldots \theta_{m-1}]^T$, and $\epsilon \sim \mathcal{N}(0, \sigma^2)$  with the variance $\sigma^2$  calculated from the vector of residuals, $\vec{r} = \vec{X}\bm{\theta}$.
Since the mean of residuals is zero in linear regression, we can compute the residual variance as:
$\sigma^2 = \frac{\vec{r}^T\vec{r}}{N} = \frac{\bm{\theta}^T\vec{X}^T\vec{X}\bm{\theta}}{N}$,
where $N$, $\bm{\theta}$, and $\vec{X}^T\vec{X}$ are already calculated in the training phase. 

\hl{
\paragraph{\bf Database perspective.}
We compute the aggregates for training and prediction using one query over the generalized cofactor ring:
}
\begin{lstlisting}[language=SQL, mathescape, columns=fullflexible]
    SELECT$\;$SUM($\lift\textsubscript{con}\hspace{-0.1em}$(X$\textsubscript{1}\hspace{-0.1em}$)$\,$* $\ldots$ *$\,\lift\textsubscript{con}$(X$\textsubscript{m}\hspace{-0.1em}$))$\;$FROM$\;$X$\textsubscript{dataset}$
\end{lstlisting}
\hl{
where $\liftcon$ is the lifting function for the generalized cofactor ring (cf. Section~\ref{sec:linear_regression}). 
Here, we assume that all attributes are continuous. 
When an attribute $X_i$ is categorical, we use the lifting function $\liftcat$ to map $X_i$-categories into ring values. 
In the presence of categorical attributes, the computed aggregate is a compact representation of the cofactor matrix computed over the one-hot encoded dataset.

We utilize user-defined functions to unpack the computed aggregate into a real-valued matrix, learn parameters $\bm{\theta}$, and compute variance $\sigma^2$.  
We generate predictions for a given test dataset as:
}
\begin{lstlisting}[language=SQL, mathescape, columns=fullflexible]
    SELECT$\;$$(\theta\textsubscript{1}\,$*$\,$X$\textsubscript{1}\,$+$\,\ldots\,$+$\,\theta\textsubscript{m-1}\,$*$\,$X$\textsubscript{m-1}\,$+$\,\epsilon)$ AS prediction FROM$\;$X$\textsubscript{test}$
\end{lstlisting}
\hl{
where $\epsilon$ is a sample from $\mathcal{N}(0, \sigma^2)$ computed using the Box-Muller transform: $\epsilon={\sqrt {-2\ln U_{1}}}\cos(2\pi U_{2}) \cdot \sigma$, where $U_1$ and $U_2$ are independent samples from the uniform distribution.
}

\subsection{Gaussian Discriminant Analysis}\label{sec:learning_classification}

We consider Gaussian discriminant analysis (GDA) for categorical data imputation. It is a type of generative classifier that models the distribution of the input features for each class as a multivariate Gaussian distribution. 
To classify a new instance, GDA estimates the posterior probability of each class given the input features using Bayes' rule and chooses the class with the highest probability. 

We focus here on Linear Discriminant Analysis (LDA), a variant of GDA where all classes share the same covariance matrix, thus yielding linear decision boundaries among classes.
The following discussion can serve as a blueprint for other types of classifiers such as Quadratic Discriminant Analysis, another variant of GDA that uses class-specific covariance matrices, and Na\"{i}ve Bayes classifiers.


\paragraph{\bf Setup.}
Consider a training dataset $\vec{X}$ consisting of $N$ training examples $(\vec{x}_1, y_1), \ldots, (\vec{x}_N, y_N)$, where the features $\vec{x}_i \in \mathbb{R}^{m}$ and the targets $y_i$ take on values from a set of classes $\{ 1, \ldots, C \}$\footnote{Without loss of generality, we assume that classes are encoded as integers.}.
GDA assumes the class-conditional densities are normally distributed:
$$\Pr(\vec{x} \,|\, y = c) = \frac{1}{(2\pi)^{m/2} |{\bm\Sigma}|^{1/2}}\exp\left (-{\frac{1}{2}(\vec{x} - {\bm\mu}_c)^T{\bm\Sigma}^{-1}(\vec{x} - {\bm\mu}_c)} \right )$$
where $m$ is the dimension of the features, ${\bm \mu}_c$ is the class-specific mean vector, and ${\bm \Sigma}$ is the shared covariance matrix for all the classes. 
Using Bayes' rule, we can compute the class posterior as:
$$\Pr(y = c \,|\, \vec{x}) = \frac{\Pr(\vec{x} \,|\, y = c) \Pr(y=c)}{\sum_{k=1}^C \Pr(\vec{x} \,|\, y = k) \Pr(y=k)}$$

We then classify a sample $\vec{x}$ into class: $\argmax_{c} \Pr (y=c \,|\, \vec{x})$.

\paragraph{\bf Training.}
We estimate the class-specific mean and shared covariance matrix from the training data.
Let $\vec{1}_{y_{i} = c}$ denote an indicator function that returns 1 if the $i$-th training example belongs to class $c$ and 0 otherwise. 
Let $\pi_c$ be the prior $\Pr(y = c)$ for class $c$.
To simplify notation, let ${\bm \theta}$ denote the parameters $({\bm\mu}_1, \ldots, {\bm\mu_C}, {\bm\Sigma}, \pi_1, \ldots, \pi_C)$.

The likelihood of data is given as:
$$L({\bm \theta}) = \Pr(\vec{X}; {\bm\theta}) = \prod_{i=1}^{N}\prod_{c=1}^{C} \Pr(\vec{x}_i | y_i = c )^{\vec{1}_{y_{i} = c}} \Pr(y_i = c)^{\vec{1}_{y_{i} = c}}$$

By maximizing $L$ with respect to the parameters ${\bm\theta}$, we find the maximum likelihood estimate of the parameters as:
\begin{align}
&
\pi_c = \frac{\sum_{i=1}^{N}{\bm 1}_{y_i = c}}{N} 
\qquad
{\bm \mu}_c = \frac{\sum_{i=1}^{N}{\bm 1}_{y_i = c}\vec{x}_i}{\sum_{i=1}^{N}{\bm 1}_{y_i = c}} \\
&\label{eq:lda_covariance}
{\bm \Sigma} = \frac{1}{N} \sum_{c=1}^{C}\sum_{i=1}^{N} {\bm 1}_{y_i = c}(\vec{x}_i - {\bm \mu}_c)(\vec{x}_i - {\bm \mu}_c)^T 
\end{align}

The class prior $\pi_c$ is the proportion of training examples that belong to the class $c$.
The class-specific mean vector ${\bm\mu}_c$ is the mean of the features of the class $c$.
The shared covariance matrix ${\bm\Sigma}_c$ is the weighted average of the covariance matrix of every class.

\paragraph{\bf Prediction.}
We can classify a sample $\vec{x}$ by finding the class $c$ that maximizes the class posterior $\Pr (y=c \,|\, \vec{x})$. After maximizing the log-posterior and dropping terms common to all classes, we obtain the classification function:
$$
f(\vec{x}) = \argmax_c\ \ln\pi_c - \frac{1}{2}(\vec{x} - {\bm\mu}_c)^T{\bm\Sigma}^{-1}(\vec{x} - {\bm\mu}_c)
$$
We can simplify $f$ by expanding the second argmax term, exploiting the symmetry of $\bm\Sigma$, and dropping terms common to all classes:
\begin{equation}\label{eq:classification_fn}
    f(\vec{x}) = \argmax_c\ \vec{a}_c^T \vec{x} + b_c
\end{equation}
where $\vec{a}_c = {\bm\Sigma}^{-1}{\bm\mu}_c$ and $b_c = \ln\pi_c - \frac{1}{2}{\bm\mu}_c^T{\bm\Sigma}^{-1}{\bm\mu}_c$.

\nop{
We can vectorize the computation of the $\argmax$ term for all classes. 
Let $\vec{M}_{m \times C} = [{\bm\mu}_1 \ldots {\bm\mu}_C]$ be the matrix consisting of class-specific mean vectors,
$\vec{A}_{C \times m}$ be the solution of the system of linear equations $\vec{A} {\bm \Sigma} = \vec{M}^T$, and $\vec{b}_{m \times 1} = [\ln\pi_1 \ldots \ln\pi_C]^T - \diag(\vec{A}\vec{M})$. Then, $\vec{A}\vec{x} + \vec{b}$ returns a $(C\times 1)$-vector with the score for each class. The index of the maximum score in the vector is the predicted class.
}

\paragraph{\bf Database Perspective.}
\nop{
Consider a training dataset defined by a join query with feature attributes $X_1, \ldots, X_m$ and a target attribute $Y$. 
The dataset has $N$ examples arranged into an $N \times (m+1)$ matrix $\vec{X}$. 
Let $N_c$ be the number of examples with class $c$:
$N_c = \sum_{i=1}^{N}{\bm 1}_{y_i = c}$.
}
We next show how to compute the aggregates needed to estimate the LDA parameters: $\pi_c$, ${\bm\mu}_c$, and $\Sigma$.
Let $N_c$ denote the number of training examples with class $c$, that is,
$N_c = \sum_{i=1}^{N}{\bm 1}_{y_i = c}$.
The \texttt{SUM(1)} aggregate counts the number of examples in the training dataset,
while \texttt{SUM(1)\;group\;by\;Y} counts the number of examples per class.
These aggregates suffice to calculate the prior $\pi_c = \frac{N_c}{N}$ for each class $c$.
To compute the class-specific mean vectors ${\bm\mu}_c$, 
we also need a batch of aggregates \texttt{SUM(X\textsubscript{i\hspace{-0.1em}})\;group\;by\;Y} for each feature attribute $X_i$, where $i\in[m]$.

To compute the shared covariance matrix,
we first rewrite the expression from Equation~\eqref{eq:lda_covariance} considering that 
$\sum_{i=1}^{N}{\bm 1}_{y_i=c}\vec{x}_i = N_c {\bm\mu}_c$:
\begin{align*}
    {\bm\Sigma} &= \frac{1}{N} \sum_{c=1}^{C}\sum_{i=1}^{N} {\bm 1}_{y_i=c} \vec{x}_i\vec{x}_i^T - \frac{1}{N}\sum_{c=1}^{C} N_c {\bm\mu}_c{\bm\mu}_c^T 
\end{align*}
The first term sums up the class-specific cofactor matrices computed over disjoint subsets of the training datasets.
This summation computes the normalized cofactor matrix 
$\frac{1}{N}(\vec{X}_{\hspace{0.1em}\text{-}Y})^T\vec{X}_{\hspace{0.1em}\text{-}Y}$,
where $\vec{X}_{\hspace{0.1em}\text{-}Y}$ is the projection of $\vec{X}$ without the target attribute $Y$.

We can compute all the required aggregates using one database aggregate query over the generalized covariance ring:

\begin{lstlisting}[language=SQL, mathescape, columns=fullflexible]
    SELECT$\;$SUM($\lift\textsubscript{con}\hspace{-0.1em}$(X$\textsubscript{1}\hspace{-0.1em}$)$\,$* $\ldots$ *$\,\lift\textsubscript{con}\hspace{-0.1em}$(X$\textsubscript{m}$)$\,$*$\,\lift\textsubscript{cat}$(Y))$\;$FROM$\;$X$\textsubscript{dataset}$
\end{lstlisting}
where $\liftcon$ and $\liftcat$ are the lifting functions for the generalized covariance ring.
Here, we assume that all input features are continuous.
The query returns a triple $(N, \vec{s}, \vec{Q})$ of aggregates.
The matrix $\vec{Q}$ of size $(m+1)\times(m+1)$ encodes the following aggregates as relations (we omit the symmetric lower part of $\vec{Q}$):
\begin{align*}
    \begin{bmatrix}
        ~{\color{blue!70!black}\texttt{SUM(X\textsubscript{1}*X\textsubscript{1})}} & {\color{blue!70!black}\cdots} & {\color{blue!70!black} \texttt{SUM(X\textsubscript{1}*X\textsubscript{m})}} & {\color{red} \texttt{SUM(X\textsubscript{1})\;group\;by\;Y}}~ \\
        & {\color{blue!70!black}\ddots} & {\color{blue!70!black}\vdots} & {\color{red}\vdots} \\
        & & {\color{blue!70!black}\texttt{SUM(X\textsubscript{m}*X\textsubscript{m})}} & {\color{red}\texttt{SUM(X\textsubscript{m})\;group\;by\;Y}} \\
        & & & {\color{green!50!black}\;\,\texttt{SUM(1)\;group\;by\;Y}} \\
    \end{bmatrix}
\end{align*}

Using the computed triple of aggregates, we can now calculate the maximum likelihood estimates of the LDA parameters as:
\begin{align*}
    \ \    
    N_c &= {\color{green!50!black}\vec{Q}_{(m+1,m+1)}}(c)
    &{\bm\mu}_c = \frac{[ {\color{red} \vec{Q}_{(1,m+1)}}(c) \ \cdots\  {\color{red}\vec{Q}_{(m,m+1)}}(c)]^T}{N_c} \\
    \pi_c &= \frac{N_c}{N} 
    &{\bm\Sigma} = \frac{1}{N}{\color{blue!70!black}\vec{Q}_{(1\ldots m,\, 1\ldots m)}} - \frac{1}{N}\sum_{c=1}^C N_c{\bm\mu}_c{\bm\mu}_c^T
\end{align*}
where $\color{blue!70!black}\vec{Q}_{(1\ldots m,\, 1\ldots m)}$ is the upper-left submatrix of $\vec{Q}$ of size $m \times m$.

When a feature attribute $X_i$ is categorical, we use $\liftcat\texttt{(X\textsubscript{i})}$ in the query computing the cofactor aggregate.
A user-defined function extracts the computed parameters into real-valued matrices and evaluates the classification function from Equation~\eqref{eq:classification_fn}. 

{



}

\section{MICE with Computation Sharing}\label{sec:indb-imputation}

The MICE algorithm iteratively trains models one after the other. During each iteration, for each attribute with missing data, the algorithm trains a model over the subset of data where the target attribute is not missing. The MICE algorithm from Section~\ref{sec:background} retrains models from scratch, discarding previous computations.
We next present our improvements over this approach.

\begin{figure*}[t]
	\begin{minipage}[t]{0.41\textwidth}
		\centering
		\includegraphics[width=\textwidth]{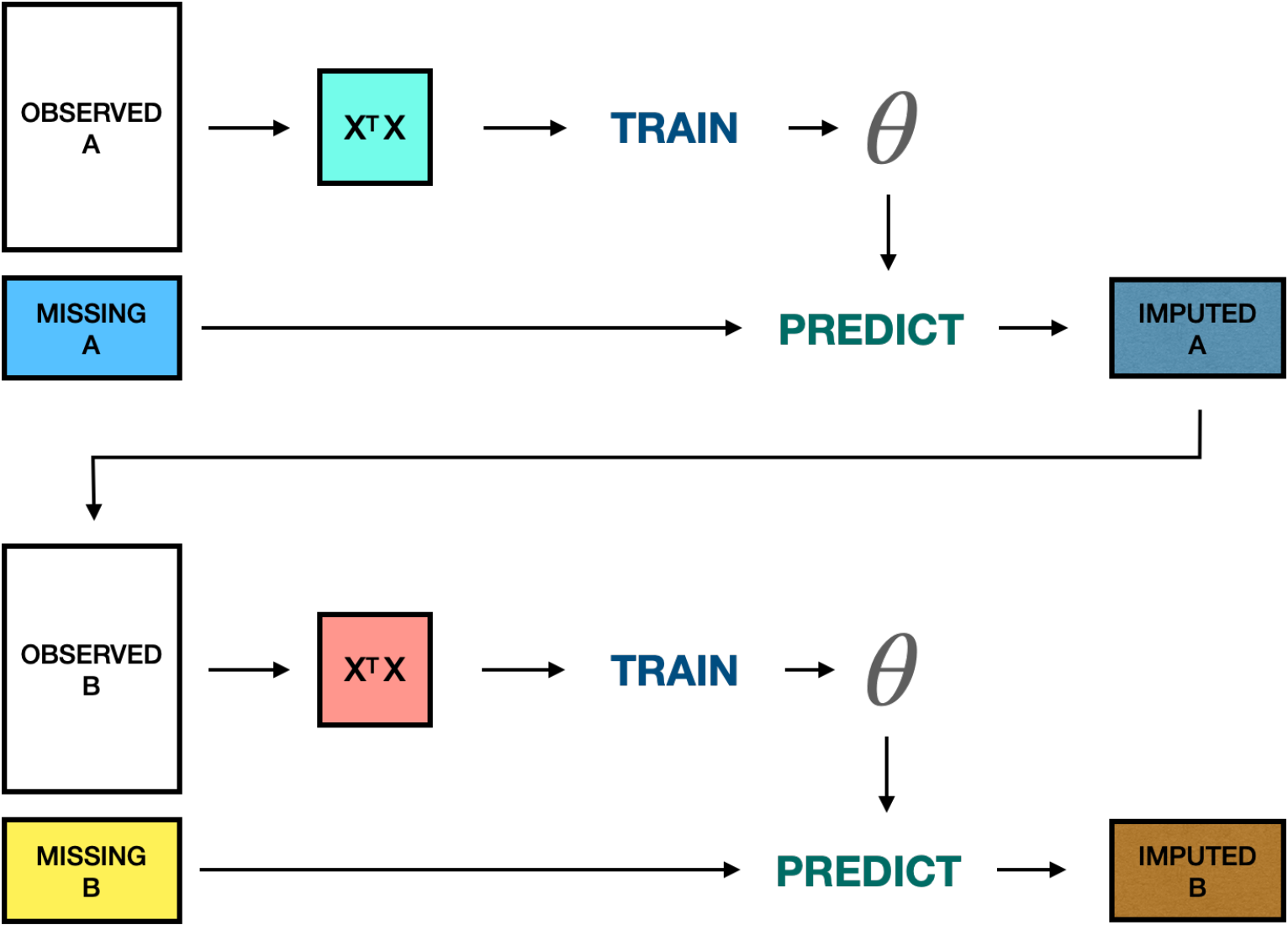}
		\vspace{2pt}
		\subcaption{MICE with in-database ML}
		\label{fig:mice_unop}
	\end{minipage}
	\hfill
	\begin{minipage}[t]{0.51\textwidth}
		\centering
		\includegraphics[width=\textwidth]{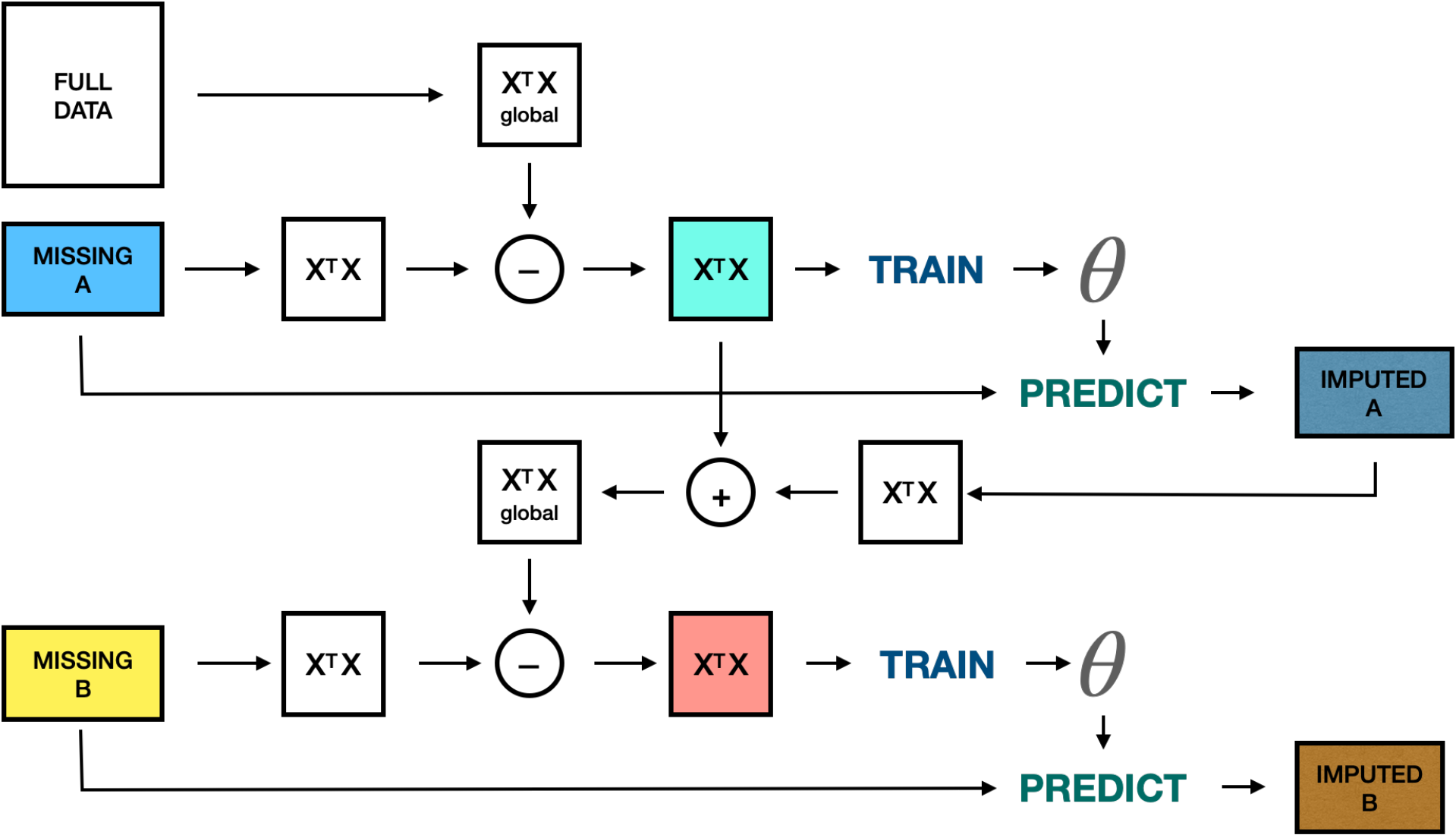}
		\vspace{2pt}
		\subcaption{MICE with in-database ML and computation sharing}
		\label{fig:mice_sharing}
	\end{minipage}
	\caption{
		Two improvements of the MICE algorithm. Colored blocks with the same color in both figures contain the same data.
	}
	\label{fig:mice_imp}
\end{figure*}

{\bf In-Database ML.}
We start by incorporating in-database learning into the MICE algorithm. 
We opt for stochastic linear regression and linear discriminant analysis as the models for imputing continuous and categorical data.
\hl{ 
These models are efficiently trainable and compute the cofactor matrix $\vec{X}^T\vec{X}$ during training, allowing for further optimizations.
}
Figure~\ref{fig:mice_unop} visualizes this approach. 
After the initial imputation, for the attribute $A$ with missing data, 
we compute the cofactor matrix over the subset of records with observed (non-missing) $A$ values, train a model, and impute the missing $A$ values. The same steps are repeated for other incomplete attributes. 

This approach benefits from faster training of individual models 
but still computes the cofactor matrix over the observed data for each incomplete attribute, in each iteration.  
When the fraction of missing values is low, as often the case in practice, recomputing the cofactor matrix due to small changes is unnecessarily expensive.

{\bf Incremental Maintenance of Cofactor Matrices.}
We can improve the previous approach using incremental computation. 
We precompute the cofactor matrix over the observed data for each incomplete attribute once.
When new imputations are created, we maintain each cofactor matrix by adjusting the contribution of the affected records before and after the imputation. 
We exploit here that the cofactor aggregates are ring values supporting $+$ and $-$.

The drawback of this approach is that it stores and maintains multiple cofactor matrices. 
Every time imputed values are updated for one attribute, each of the cofactor matrices might be affected due to the overlap among the datasets used for their computation.

{\bf Shared Computation of Cofactor Matrices.}
\hl{
We next present two optimizations designed to boost the performance of MICE by enabling computation sharing across iteration. 
The first optimization targets datasets with low missing rates, while the second optimization applies to both low and high missing rate datasets.  

{\em (1) Shared Computation with Low Missing Rates.}
}
In practice, the observed data is often much larger than the missing data. We leverage this observation to compute one global cofactor matrix over the entire (initially imputed) data once, 
and use it to derive the cofactor matrix for each attribute on-the-fly, by scanning only the missing data. 
This approach effectively moves the expensive computation over the observed data outside the iteration loop.

Algorithm~\ref{alg:mice_sharing_v2} shows our MICE algorithm with computation sharing. After the initial imputation, we compute the global cofactor matrix over the imputed dataset $\tilde{\vec{X}}$ (Line~2). 
To compute the cofactor matrix for an attribute $X_i$, 
we remove the contribution of the records with missing $X_i$-values, $\tilde{\vec{X}}_{i=\mathit{miss}}$, from the global cofactor matrix since these records are not used for training (Lines~5-6).
After training the model and imputing new values,
we update the global matrix to account for the new imputations (Lines~9-10),
making it ready for the next iteration. 
Figure~\ref{fig:mice_sharing} visualizes this approach.
The colored blocks denote the same data in the two approaches, with and without sharing the cofactor matrix computation. 

\hl{
{\em (2) Shared Computation with Data Partitioning.}
We can further accelerate the cofactor matrix computation by partitioning the data into partitions based on the number of missing values in each record. 
We start with two observations.
First, the records containing no missing values are part of every training dataset and contribute equally to the cofactor matrix computed in each iteration. Thus, we can form a partition of such records, precompute their contribution once, and reuse it in every iteration.
Second, the records containing missing values in all incomplete attributes are not part of any training dataset, thus we can skip such records during model training and only impute them at the end of each iteration. 

For datasets with low missing rates, 
we want to ensure fast access to (small) incomplete data, $\tilde{\vec{X}}_{i=\mathit{miss}}$, needed by Algorithm~\ref{alg:mice_sharing_v2}.
Before starting the imputation, we split each dataset into four partitions: 
one stores records without missing values, 
one stores records with only missing values,
one stores records with exactly one missing value, and
one stores records with at least two missing values. 
We further recursively partition the third partition containing records with one missing values into subpartitions, one for each incomplete attribute.
Then, accessing the records with missing values in a given attribute requires scanning two partitions: 
the subpartition of the given attribute and the third (`overflow') partition.
When the fraction of missing values is low, both of these partitions tend to be small, allowing fast access to the needed records. 

For datasets with high missing rates, 
we want to ensure fast access to (small) observed data, $\tilde{\vec{X}}_{i=\mathit{obs}}$, to speed up the model training in Algorithm~\ref{alg:mice_v2}.
The partitioning strategy in this case is the complete opposite of that in the case with low missing rates: 
partitioning uses the same criteria but based on the number of observed (not missing) values of incomplete attributes in each record.
The partition containing records with only observed values is used to precompute a partial cofactor aggregate once, outside the iteration loop. 
The rest of the training dataset for an incomplete attribute contains records from two partitions: one subpartition storing records with exactly one observed value for the given attribute and one partition storing records with at least two observed values. 
The model training in each iteration now needs to scan only these two partitions, often much smaller than the entire dataset when its missing rate is high.
}

\nop{
\hl{
While this approach is useful in the case of low missing rate due to the faster computation of the cofactor matrix over this portion of the dataset, in the case of high missing rate this is not the case. In this scenario the training set of each model is significantly reduced, therefore is faster to recompute the cofactor matrix over the training set. Optimizations which can be introduced in this scenario are computing once and sharing the cofactor matrix over the tuples without missing values and ignore tuples which consist of all missing values. }
}

\begin{algorithm}[t]
	\caption{MICE with computation sharing} \label{alg:mice_sharing_v2}
	\SetAlgoLined
	\SetKwInput{Input}{Input\hspace{0.7em}}
    \SetKwInput{Output}{Output}
	\SetKwProg{Fn}{Function}{:}{}
	\Input{
		\begin{tabular}[t]{ll@{}}
			$\vec{X}$ & incomplete dataset with attributes $X_1, ..., X_m$ \\
			$\mathit{mattrs}$ & indices of incomplete attributes
		\end{tabular}
	}
	\Output{
		\begin{tabular}[t]{ll}
			$\tilde{\vec{X}}$ \hspace{1.6em} & imputed dataset
		\end{tabular}
	}
	{
		$\tilde{\vec{X}} \leftarrow \vec{X}$ with initial imputations for all missing values\\
		$\vec{C} \leftarrow \tilde{\vec{X}}^{\text{T}} \tilde{\vec{X}}$\\
		\Repeat{ $\mathit{stopping\_condition}$}{
			\ForEach{$i \in \mathit{mattrs}$}{
				$\Delta\vec{C} \leftarrow (\tilde{\vec{X}}_{i=\mathit{miss}})^{\text{T}}\, \tilde{\vec{X}}_{i=\mathit{miss}}$\\
				$\vec{C}_{\mathit{train}} \leftarrow \vec{C} - \Delta\vec{C}$\\
				$\bm{\theta} \leftarrow$ \textsc{Train}($\mathit{cofactor} = \vec{C}_{\mathit{train}}$,\, $\mathit{target} = X_i$) \\
				$\tilde{\vec{X}} \leftarrow \textsc{Predict}(\mathit{data} = \tilde{\vec{X}}_{i=\mathit{miss}},\, \mathit{target} = X_i, \mathit{model} = \bm{\theta})$ \\
				$\Delta\vec{C} \leftarrow (\tilde{\vec{X}}_{i=\mathit{miss}})^{\text{T}}\, \tilde{\vec{X}}_{i=\mathit{miss}}$\\
				$\vec{C} \leftarrow \vec{C}_{\mathit{train}} + \Delta\vec{C}$\\
			}
		}
	}
\end{algorithm}

\nop{
\begin{algorithm}[t]
	\SetAlgoLined
	\small
	\SetKwProg{Fn}{Function}{:}{}
	\textbf{Inputs:} 
	$X$;  \tcp{dataset}\
	$I$;  \tcp{MICE iteration}\
	$Ioops$;  \tcp{MICE iterations}\
	$c $; \tcp{column}\
	$cols $; \tcp{columns with missing values}\
	$\theta $; \tcp{model parameters}\
	$X_{c=miss}$;  \tcp{tuples of the dataset originally missing in column $c$}\
	\Fn{MICE()}
	{
		cofactor = $X^T \cdot X$\;
		\label{alg_2:cofactor}
		\ForEach{$l$ $\in$ $loops$}
		{
			\ForEach{$c$ $\in$ $cols$}
			{
				cofactor\_delta = $X_{c=miss}^T \cdot X_{c=miss}$\;
				train\_cofactor = cofactor - cofactor\_delta\;
				\label{alg_2:train_cofactor}
				$\theta$ = train\_model(train\_cofactor, c)\;
				$X_{c=miss}$ = generate\_predictions($X_{c=miss}$, $\theta$)\;
				cofactor\_delta = $X_{c=miss}^T \cdot X_{c=miss}$\;
				cofactor = train\_cofactor + cofactor\_delta\;
				\label{alg_2:new_cofactor}
			}
		}
	}
	\caption{MICE with sharing of computations}
	\label{alg:MICE_sharing}
\end{algorithm}
}

\section{Implementation}
\label{sec:implementation}
We next discuss how to implement our data imputation methods in existing DBMSs.
We opt for an in-database implementation instead of building a specialized tool for several reasons:


\begin{itemize}
	\item 
	Providing a solution integrated with widely-used DBMSs facilitates its adoption, eliminating the need for adding yet another tool to already complex data pipelines. 
	
	\item 
	DBMSs are mature systems with robust mechanisms for handling large data and recovering from failures, in contrast to data imputation implementations in tools such as R.
	
	\item
	DBMSs offer highly-optimized techniques for query evaluation, including parallel execution, allowing us to focus on more high-level aspects of the imputation process.
\end{itemize}

We implemented our imputation methods in PostgreSQL and DuckDB, two open-source database systems with row-oriented and column-oriented storage models, respectively.  
Our approach can be implemented in any other database system that supports defining custom data types of variable size and aggregate functions operating over values of these types. 
Our implementation assumes that categorical values are encoded as integers; if they are not, we can map categorical values to integers in a preprocessing step.

\subsection{In-Database ML Implementation}
We provide libraries in PostgreSQL and DuckDB for in-database training of ridge linear regression, stochastic linear regression, and LDA models, and for generating predictions under each model. 
The core library component is the implementation of the generalized cofactor ring. 
We refer to this data structure as \textsc{Triple}.

In PostgreSQL, \textsc{Triple} is a custom data type of variable size. A \textsc{Triple} value is a struct that comprises 
an array storing the numerical aggregates over continuous attributes, followed by
an array storing the relational aggregates over continuous and categorical attributes.
The struct occupies a contiguous memory chunk with no external pointers to allow for fast allocation/de-allocation.

DuckDB, instead, offers native support for nested data types such as arrays and structs, allowing for a simpler \textsc{Triple} implementation based on using an array of numerical values and an array of key-value structs. 

In both cases, we implement the ring operations, addition, subtraction, and multiplication, over \textsc{Triple} values, as user-defined functions.
Instead of the lifting functions $\liftcon$ and $\liftcat$,
we provide a more efficient bulk version $\lift$ that takes as input a list of continuous attributes and a list of categorical attributes and maps their values to a \textsc{Triple} aggregate at once, avoiding triple multiplication. 

\begin{lstlisting}[language=SQL, mathescape, columns=fullflexible]
    SELECT$\;$SUM($\,\lift\hspace{-0.05em}$($\,$[X$\textsubscript{i\textsubscript{1}},\ldots,$X$\textsubscript{i\textsubscript{k}}$], [X$\textsubscript{j\textsubscript{1}},\ldots,$X$\textsubscript{j\textsubscript{l}}$]$\,$))$\;$FROM$\;$X$\textsubscript{dataset}$
\end{lstlisting}

DuckDB supports a faster way of computing \textsc{Triple}s via a custom aggregate operator that operates directly over attribute values:
\begin{lstlisting}[language=SQL, mathescape, columns=fullflexible]
    SELECT$\;$SUM_TRIPLE(X$\textsubscript{i\textsubscript{1}},\ldots,$X$\textsubscript{i\textsubscript{k}}$,$\,$X$\textsubscript{j\textsubscript{1}},\ldots,$X$\textsubscript{j\textsubscript{l}}$)$\;$FROM$\;$X$\textsubscript{dataset}$
\end{lstlisting}
In the implementation, the aggregate operator \texttt{\small SUM\_TRIPLE} takes as input a list of value vectors of a fixed size, together with their type (continuous or categorical), and aggregates these values to a \textsc{Triple} in bulk, rather than one record at a time. 

{\bf Factorized Computation of the Cofactor Matrix.}
When computing the cofactor matrix over joins, 
we can exploit the algebraic properties of the cofactor ring to achieve factorized evaluation, that is, compute partial \textsc{Triple} values over individual tables and then combine these triples to produce the final result.

\begin{example}\label{ex:factorized_cofactor}
	Consider relations $R(A,B)$ and $S(B,C,D)$, where $C$ is categorical and the others are continuous.
	We can compute the cofactor triple for $A$, $C$, and $D$ over the join of $R$ and $S$ as:
\begin{lstlisting}[language=SQL, mathescape, columns=fullflexible]
    SELECT$\;$SUM($\,\lift\hspace{-0.05em}$($\,$[A,D],$\,$[C]$\,$))$\;$FROM$\;$R,$\,$S$\;$WHERE$\;$R.B$\,$=$\,$S.B 
\end{lstlisting}
The factorized query computes partial triples before the join:
\begin{lstlisting}[language=SQL, mathescape, columns=fullflexible]
	SELECT$\;$SUM(t1.T$\;$*$\;$t2.T) FROM
	  (SELECT$\;$B,$\;$SUM($\lift\hspace{-0.05em}$($\,$[A],$\,$[$\;\,$]$\,$))$\;$AS$\;$T$\;$FROM$\;$R$\;$GROUP$\;$BY$\;$B)$\;$AS$\;$t1,
	  (SELECT$\;$B,$\;$SUM($\lift\hspace{-0.05em}$($\,$[D],$\,$[C]$\,$))$\;$AS$\;$T$\;$FROM$\;$S$\;$GROUP$\;$BY$\;$B)$\;$AS$\;$t2
	WHERE$\;$t1.B$\;$=$\;$t2.B
\end{lstlisting}
This optimization often pays off when the domain of $B$ is small.
\punto
\end{example}

In the current implementation, we refactor input queries like in Example~\ref{ex:factorized_cofactor} before passing them to the query optimizer. 
As future work, we aim to enable this optimization in the query optimizer.


The \textsc{Triple} value computed over the training dataset is passed to the functions for model training and prediction, along with other parameters such the index of the target attribute, learning rate, and regularization factor. 
LDA relies on LAPACK routines~\cite{Angerson:1990:LAPACK} to solve systems of linear equations and perform matrix operations, necessary in model training and prediction.

\subsection{In-Database MICE Implementation}\label{sec:mice_implementation}

\hl{
	We implemented our imputation methods as driver functions in PL/pgSQL for PostgreSQL and in C++ for DuckDB.
	We provide three functionally-equivalent implementations of the MICE algorithm:
	\begin{enumerate}
		\item \textsc{Baseline} implements the logic from Algorithm~\ref{alg:mice_v2} using the generalized cofactor ring, without any partitioning strategy;
		\item \textsc{Low} implements the shared cofactor computation from Algorithm~\ref{alg:mice_sharing_v2} using the partitioning strategy for datasets with low missing rates;
		\item \textsc{High} implements the shared cofactor computation using the partitioning strategy for datasets with high missing rates.
	\end{enumerate}

	Each implementation creates a copy of the dataset that needs to be imputed in the preprocessing step, outside the iteration loop. 
	The \textsc{Baseline} version repeatedly scans the entire copy to train models over subsets of records with complete values for different incomplete attribute. The other two versions create a partitioned copy and compute partial cofactor aggregates in the preprocessing step, as discussed in Section~\ref{sec:indb-imputation}. This partitioning can accelerate the retrieval of matching records, reducing the per-iteration cost at the expense of increasing the preprocessing cost.
}

{\bf Reducing Update Overhead.}
Both PostgreSQL and DuckDB use multi-version concurrency control (MVCC) to maintain data consistency.
When imputing new values using an \texttt{UPDATE} command, MVCC creates new versions of the affected objects. 
As the algorithm proceeds iteratively, updating imputed values leads to increasingly many obsolete versions, causing significant overheads and eventually becoming the bottleneck in our implementation.  

A potential solution is to store each attribute's imputed values in another table, re-created every time these values are updated. The main drawback of this solution, however, is that it requires joining all tables of imputed values every time a model is trained.

We reduce the update overhead with the following optimizations:
\begin{itemize}
	\item {\em Swapping columns (DuckDB)}.
	We changed the internals of DuckDB to support pointer-based column swaps between tables (70 LOC), inspired by a similar approach from prior work~\cite{Huang:2023:JoinBoost}. 
	We create a temporary table with one column containing new imputed values and then move this column in place of an existing one in the corresponding table partition. 
		
	\item {\em Recreating subpartitions (DuckDB \& PostgreSQL)}.
	The subpartitions storing records with one missing value are completely changed in each iteration. 
	In PostgreSQL, we recreate such subpartitions with new imputations but update in-place other affected partitions.
	In DuckDB, we always swap a fresh column with new imputations into affected partitions. 	

	\item {\em Enabling heap-only-tuples (PostgreSQL).}
	This optimization in PostgreSQL aims to store multiple versions of one row on the same page, reducing update overheads. 
	For partitions updated in-place (e.g., those storing records with at least two missing values), we reduce their page fill factor to $75\%$.

	\item {\em Using shorter transactions (PostgreSQL).} 
	The vacuum processes in PostgreSQL can only remove row versions that are older than any currently active transaction. Repeatedly updating the same rows in a single transaction leads to accumulating obsolete versions and degrading performance. Thus, we start a new transaction on imputing each attribute.
\end{itemize}

\section{Experiments}\label{sec:experiments}

We compare our techniques for in-database learning and data imputation in DuckDB and PostgreSQL against the following systems:
Apache SystemDS~\cite{system:SystemDS}, a data science platform for large-scale data analysis and machine learning;
Apache MADlib~\cite{system:MADlib}, a PostgreSQL library for in-database machine learning;
MindsDB~\cite{system:MindsDB}, a layer on top of database systems for training machine learning models; and five alternative imputation methods implemented in Python. 


Our experimental findings can be summarized as follows:
\begin{itemize}
	\item 
	Using the cofactor ring for computing cofactor aggregates improves the training performance by up to 6x, regardless of the dataset and DBMS. Factorized computation of the cofactor aggregates can further improve the performance up to 12x if the denormalized database is highly redundant.
	
	\item 
	When imputing a single table, our DuckDB implementation outperforms the fastest competitor, SystemDS, in terms of per-iteration cost by 86x to 346x as the rate of missing values ranges between 5\% and 80\%. The imputation time scales linearly with the number of incomplete attributes.

	\item When imputing missing values over a normalized dataset, 
	employing factorized evaluation can be faster than denormalizing the dataset before imputation 	by up to 6 times in PostgreSQL and up to 1.7 times in DuckDB. 	

	\item Our MICE implementations are competitive with or outperform state-of-the-art imputation methods in terms of imputation quality, while offering up to two orders of magnitude faster imputation, under various missing rates and patterns. 

\end{itemize}



{\bf Experimental Setup.}
We run all experiments on a server with 2 x AMD EPYC 7302 16-Core Processor, 64 threads, 512 GB RAM running Ubuntu 20.04. We use PostgreSQL 12.12 and DuckDB 0.8.1. We run each experiment 3 times with a timeout of 200 minutes and report averaged results, unless stated otherwise.


{\bf Datasets. }
We consider three real-world datasets:
(1) {\em Flight Delays and Cancellations}~\cite{dataset:flight}
	contains information about U.S. flights. It consists of 3 tables, 60M rows and 31 columns, of which 5 are categorical.
(2) {\em Retailer}~\cite{Schleich:2016:Learning} contains historical inventory data of stores at different locations. It consists of 5 tables arranged in a snowflake schema, 84M rows and 25 columns, of which 4 are categorical.
(3) {\em Taiwan's Air Quality}~\cite{dataset:air_quality} contains information about Taiwan's air quality in the years 2016-2021. It is a single table with 3.5M rows, 11 numerical columns, and 6\% missing values.


The Flight and Retailer datasets are complete, containing no missing values. These datasets serve as the foundation for our benchmarks, where we randomly remove different quantities of values to evaluate performance under various scenarios. 


\begin{figure*}[t]
	\begin{minipage}[b]{.45\textwidth}
		\centering
		\includegraphics[width=0.7\textwidth]{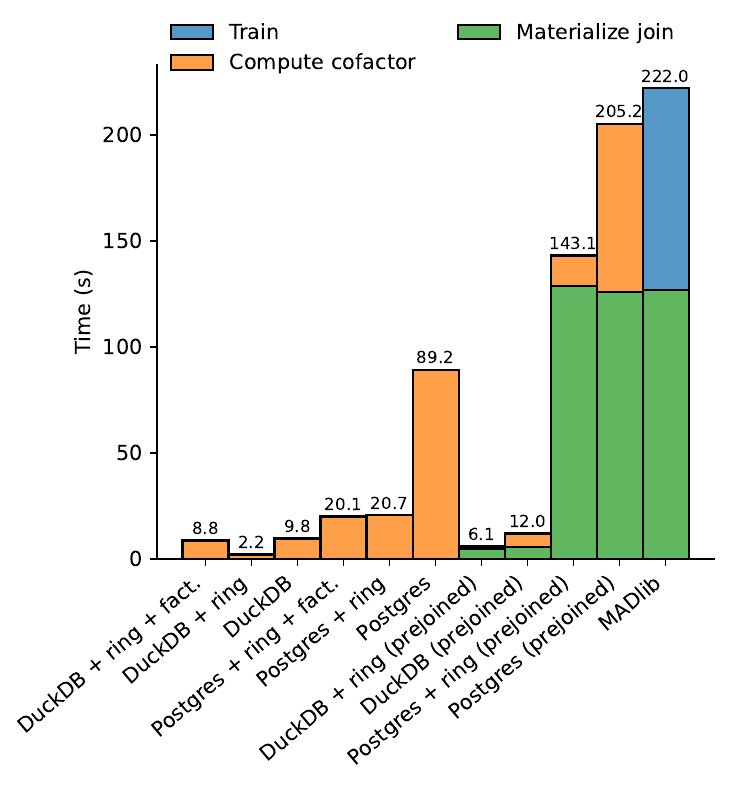}
		\subcaption{Flight dataset (cont. only)}
		\label{fig:train_flight_num}
	\end{minipage}
	\qquad
	\begin{minipage}[b]{.45\textwidth}
		\centering
		\includegraphics[width=0.7\textwidth]{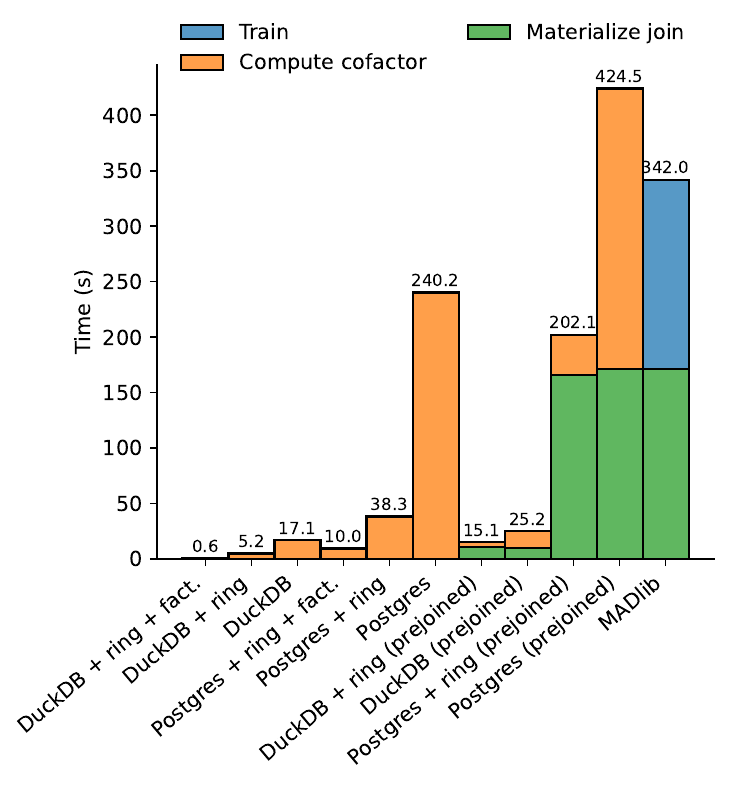}
		\subcaption{Retailer dataset (cont. only)}
		\label{fig:train_retailer_num}
	\end{minipage}
	
	\begin{minipage}[b]{.45\textwidth}
		\centering
		\includegraphics[width=0.7\textwidth]{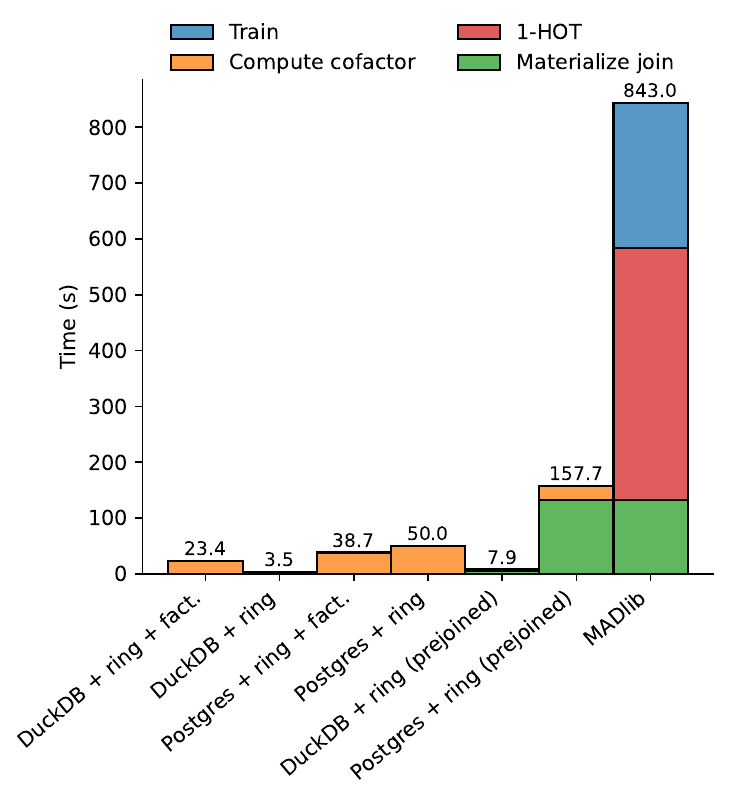}
		\subcaption{Flight dataset (cont. \& cat.)}
		\label{fig:train_flight_cat}
	\end{minipage}  
	\qquad
	\begin{minipage}[b]{.45\textwidth}
		\centering
		\includegraphics[width=0.7\textwidth]{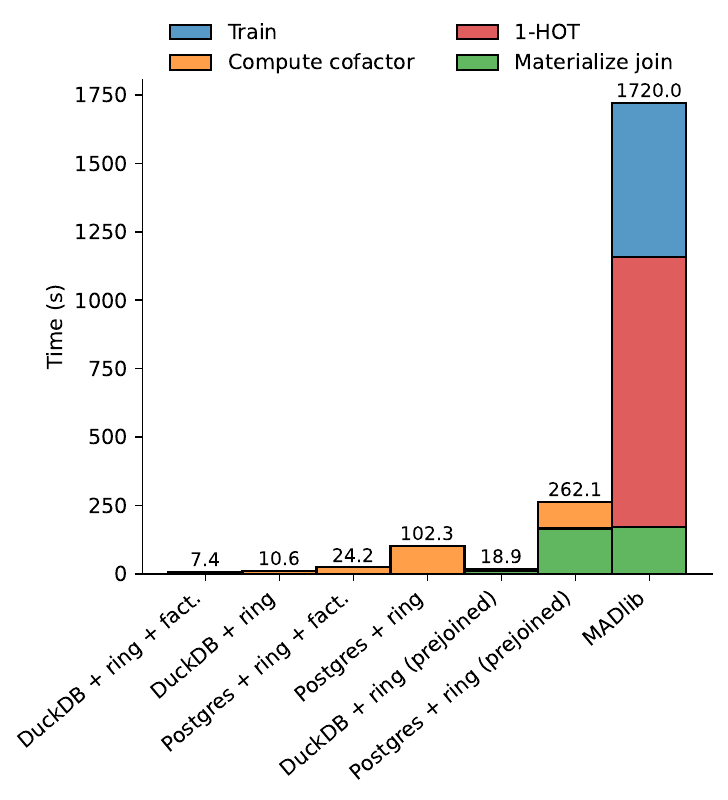}
		\subcaption{Retailer dataset (cont. \& cat.)}\label{fig:train_retailer_cat}
	\end{minipage}  
	\caption{
		Breakdown of execution time for training a linear regression model in DuckDB, PostgreSQL, and MADlib over the join of the input tables from the Retailer and Flight datasets, with continuous only and continuous + categorical attributes.
		The \texttt{ring} variants use the covariance ring, \texttt{fact} uses factorized evaluation, and \texttt{prejoined} materializes the join result first. 
	}
	\label{fig:train_exp}
\end{figure*}

\subsection{In-Database Learning}

We evaluate the performance of training a linear regression model inside DuckDB and PostgreSQL when the training dataset is formed by joining the input tables. 
We consider three different evaluation approaches for DuckDB and PostgreSQL.
The baseline approach uses standard SQL with scalar \texttt{SUM} aggregates to compute the cofactor matrix, 
the second approach adopts the cofactor ring (\texttt{ring}), and
the third approach additionally includes factorized evaluation (\texttt{ring + fact}).
We also compare our library with MADlib. We run this experiment over the Flight and Retailer datasets considering only continuous and both continuous and categorical attributes. MADlib assumes a single table is available, therefore, we need to precompute and materialize the joined relation. 
We are unable to compute the cofactor aggregates with a standard SQL query over categorical attributes, because the number of aggregates after one-hot encoding exceeds the limits in PostgreSQL and DuckDB.

Figure~\ref{fig:train_exp} shows the performance of these approaches. In the case of continuous-only attributes, the adoption of the ring structure leads to 4 to 6 times higher performance, regardless of the DBMS used \hl{due to the compact representation of the cofactor aggregates}. While the addition of categorical attributes slows down the computation, the performance improvement over MADlib is even higher, as one-hot encoding in MADlib takes 450s for  Flight and 980s for Retailer, in addition to materialization of the joined result.

Computing the cofactor aggregates using factorized evaluation affects the performance differently according to the dataset used. \hl{In DuckDB, training a model over the normalized Retailer dataset improves the performance by 8.7x with continuous-only attributes and by 1.4x with mixed attributes compared with non-factorized evaluation. 
The smaller speedup with categorical attributes is due to increased memory management pressure caused by resizing the data structures that store categorical values.}
With the Flight dataset, factorized evaluation leads to a higher runtime because joining the input tables does not produce many redundant values: the fact table already contains most of the data, with the other tables being 15\% of its size. In Retailer, joining the input tables brings more redundancy as the fact table has only 4 attributes, with the other tables being less than 1\% of its size. The computation saved by factorized evaluation is therefore more pronounced in Retailer than Flight.


\nop{
Our experiments with PostgreSQL obtain similar results. Training a model using factorized evaluation on the Retailer dataset leads to 5 times better performance regardless of the inclusion of categorical columns, as PostgreSQL aggregate functions do not use a shared state. Over the Flight dataset, instead, training a model over the normalized database and over the joined relation perform similarly, as the lower efficiency of the factorized approach is balanced by PostgreSQL's slower computation of the join result.
}


\begin{figure*}
	\centering
	\begin{minipage}[t]{0.27\textwidth}
		\centering
		\includegraphics[width=\textwidth]{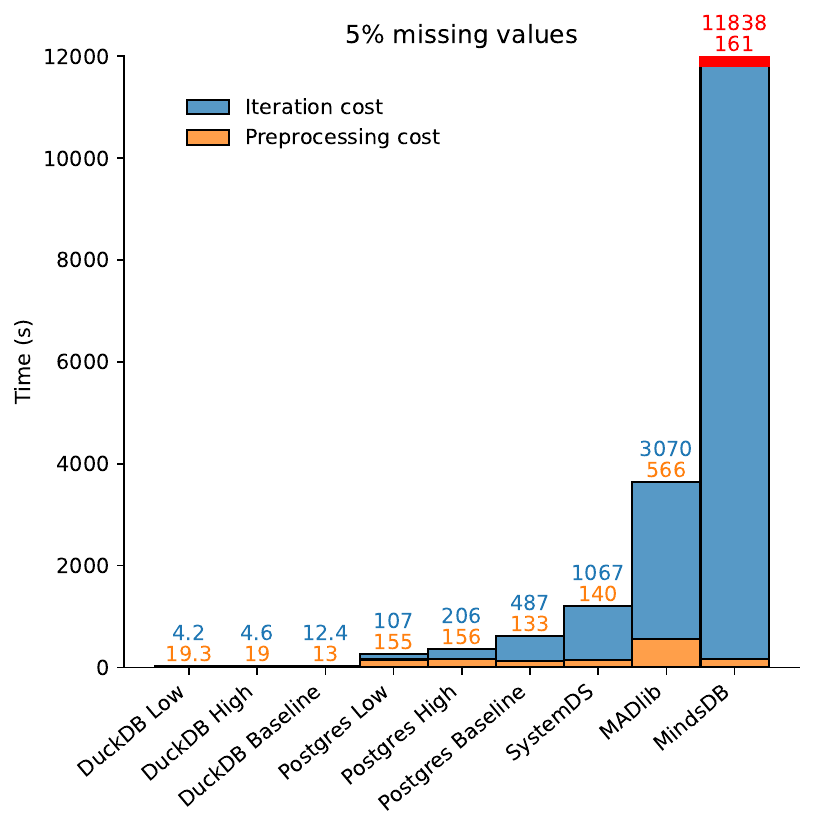}
		\subcaption{Flight 5\% missing}
		\label{fig:single_table_flight_0.05}
	\end{minipage}
	\hfill
	\begin{minipage}[t]{0.27\textwidth}
		\centering
		\includegraphics[width=\textwidth]{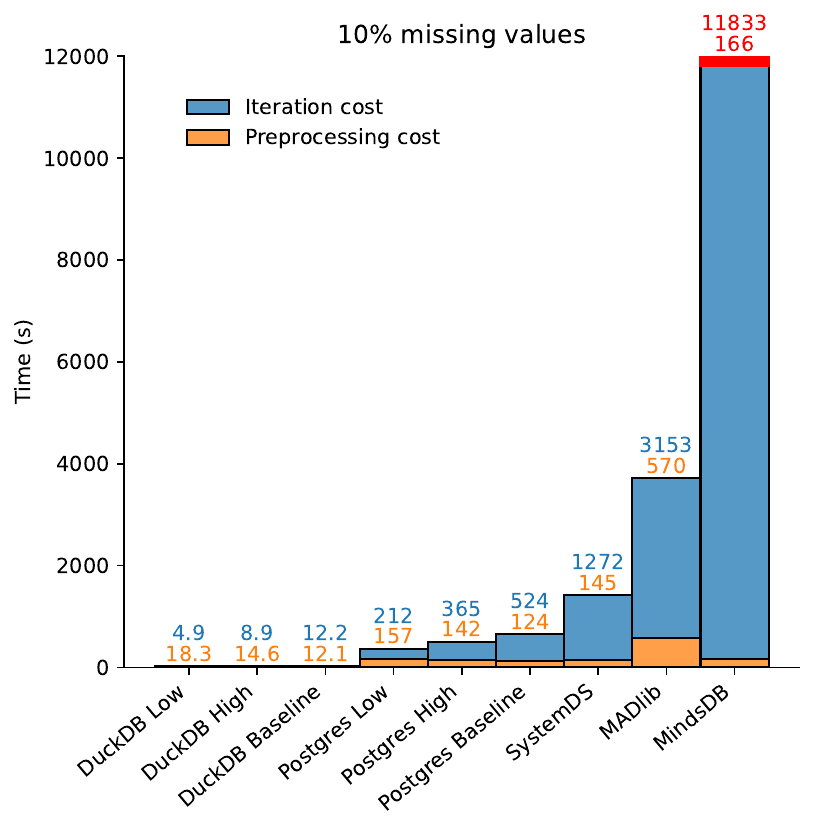}
		\subcaption{Flight 10\% missing}
		\label{fig:single_table_flight_0.1}
	\end{minipage}
	\hfill
	\begin{minipage}[t]{0.27\textwidth}
		\centering
		\includegraphics[width=\textwidth]{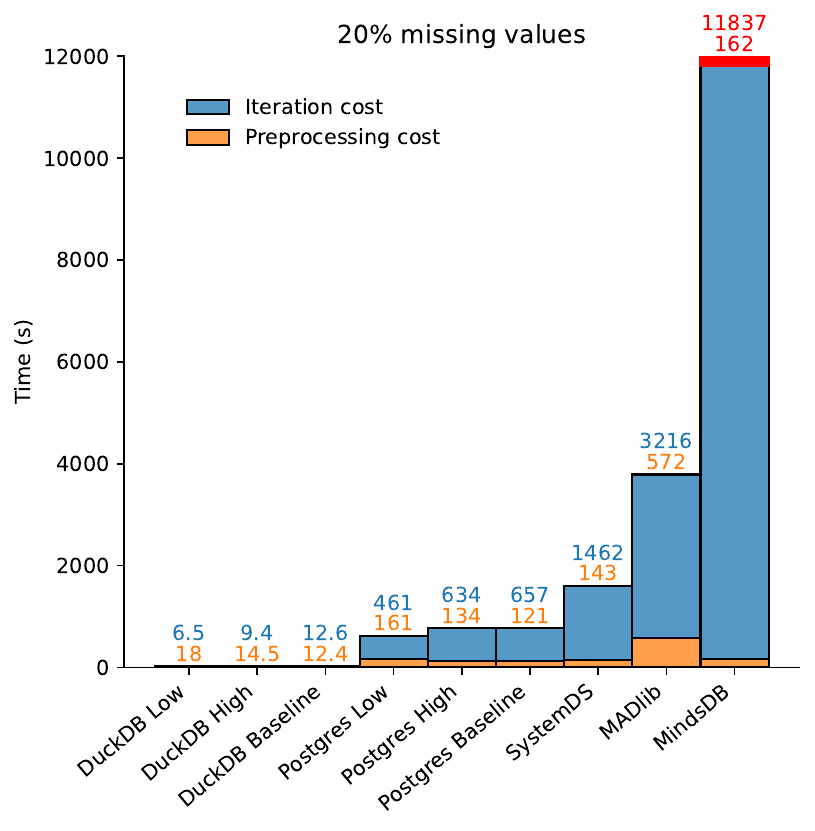}
		\subcaption{Flight 20\% missing}
		\label{fig:single_table_flight_0.2}
	\end{minipage}
	
	\begin{minipage}[t]{0.27\textwidth}
		\centering
		\includegraphics[width=\textwidth]{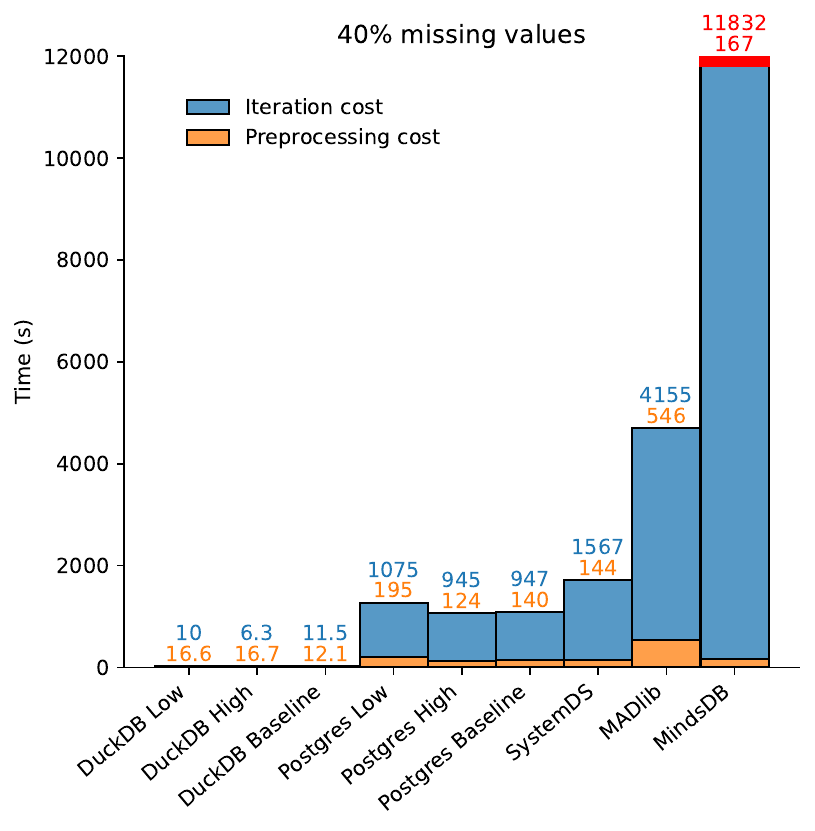}
		\subcaption{Flight 40\% missing}
		\label{fig:single_table_flight_0.4}
	\end{minipage}
	\hfill	
	\begin{minipage}[t]{0.27\textwidth}
		\centering
		\includegraphics[width=\textwidth]{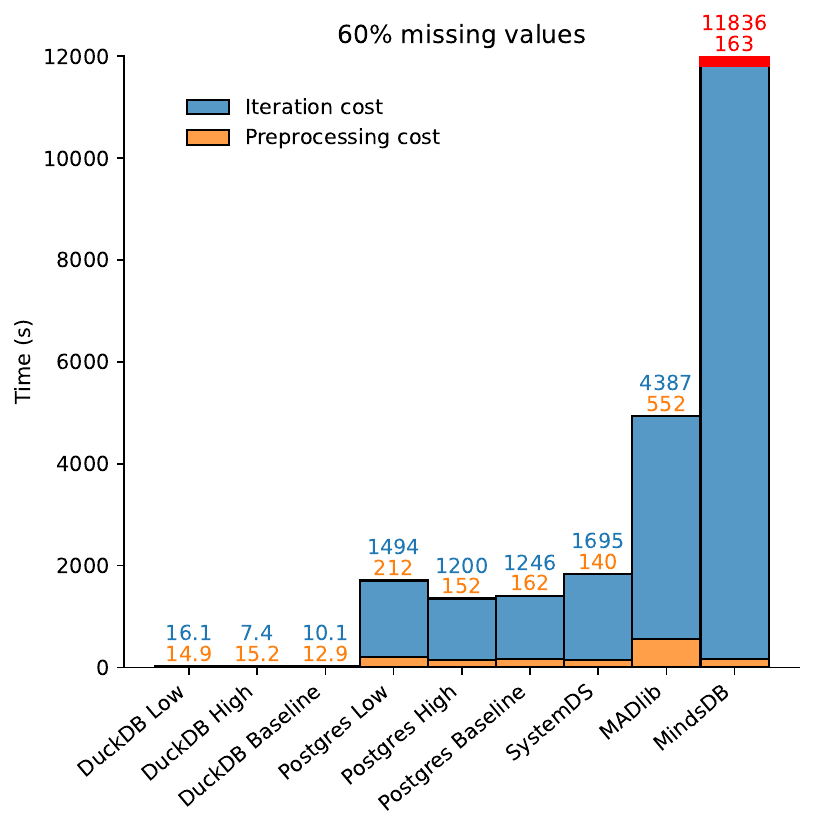}
		\subcaption{Flight 60\% missing}
		\label{fig:single_table_flight_0.6}
	\end{minipage}
	\hfill
	\begin{minipage}[t]{0.27\textwidth}
		\centering
		\includegraphics[width=\textwidth]{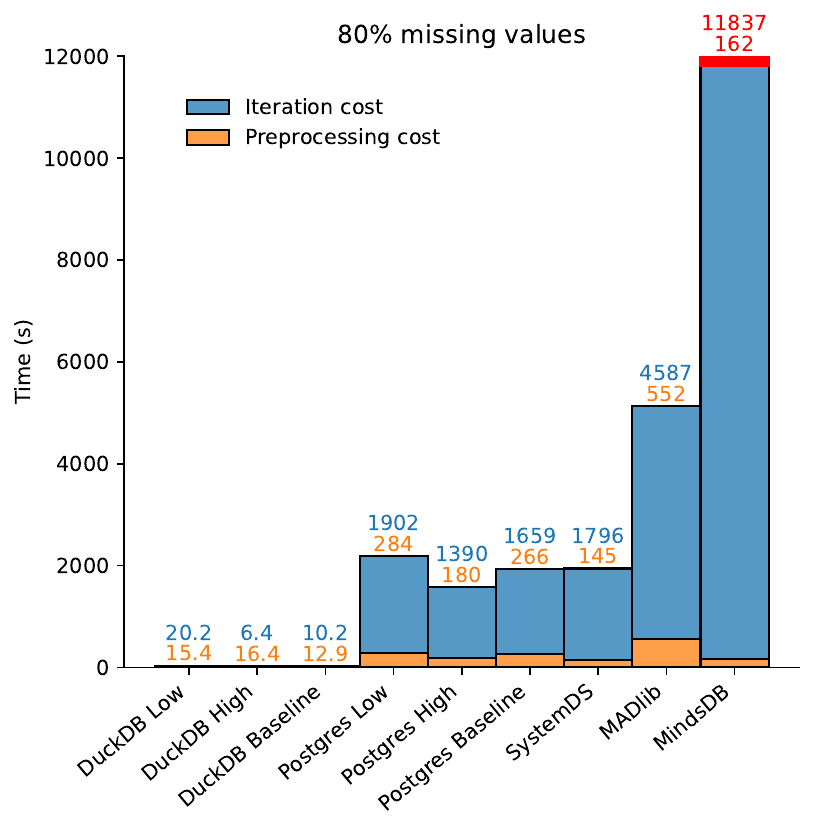}
		\subcaption{Flight 80\% missing}
		\label{fig:single_table_flight_0.8}
	\end{minipage}
		
	\begin{minipage}[t]{0.27\textwidth}
		\centering
		\includegraphics[width=\textwidth]{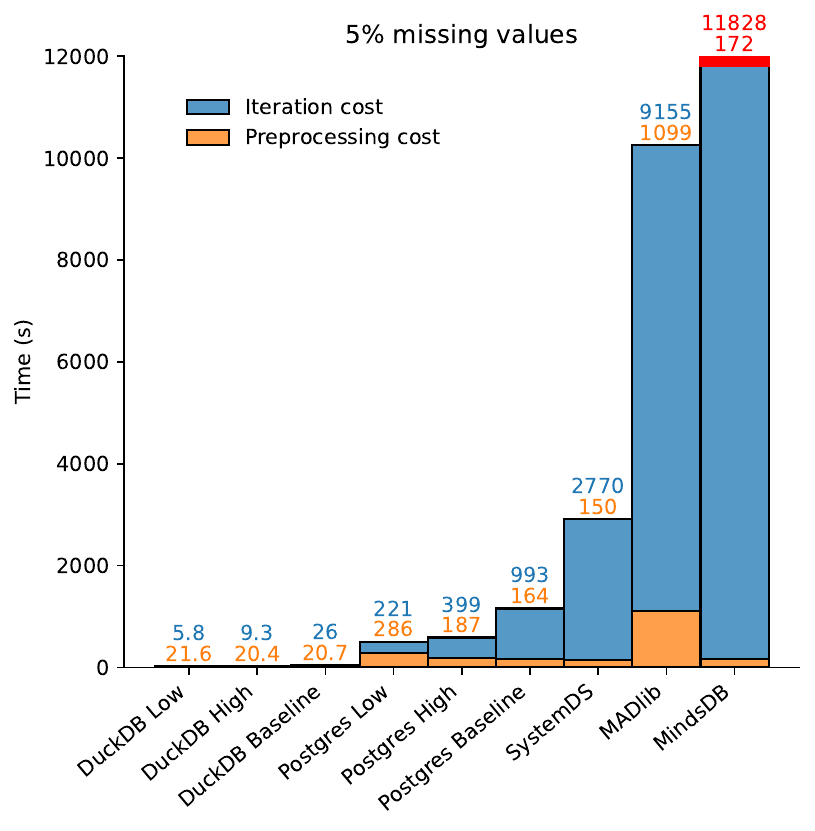}
		\subcaption{Retailer 5\% missing}
		\label{fig:single_table_retailer_0.05}
	\end{minipage}
	\hfill
	\begin{minipage}[t]{0.27\textwidth}
		\centering
		\includegraphics[width=\textwidth]{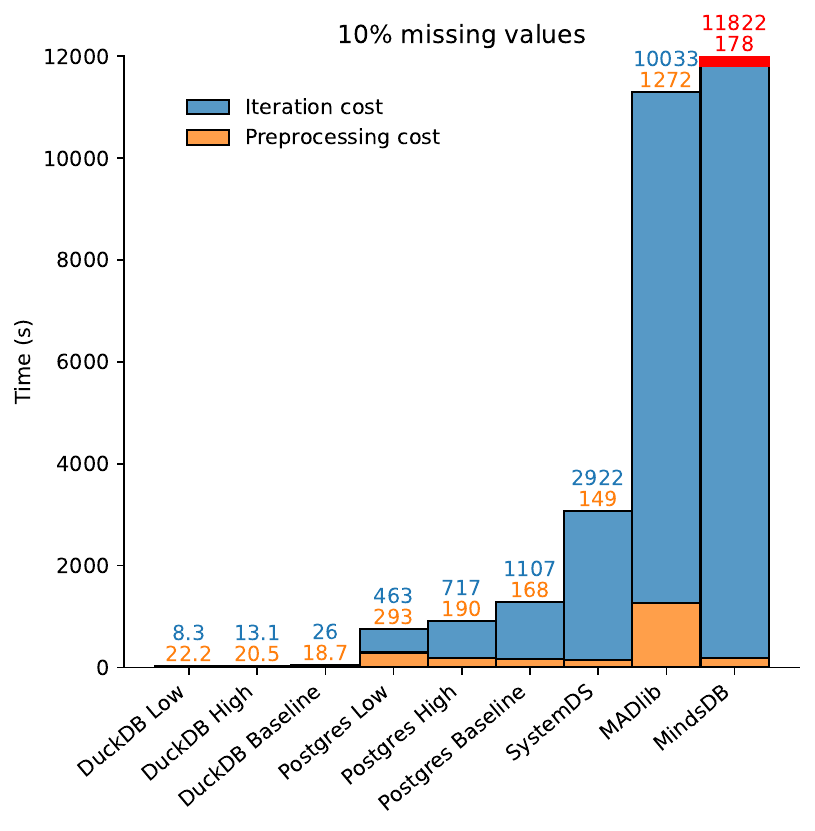}
		\subcaption{Retailer 10\% missing}
		\label{fig:single_table_retailer_0.1}
	\end{minipage}
	\hfill
	\begin{minipage}[t]{0.27\textwidth}
		\centering
		\includegraphics[width=\textwidth]{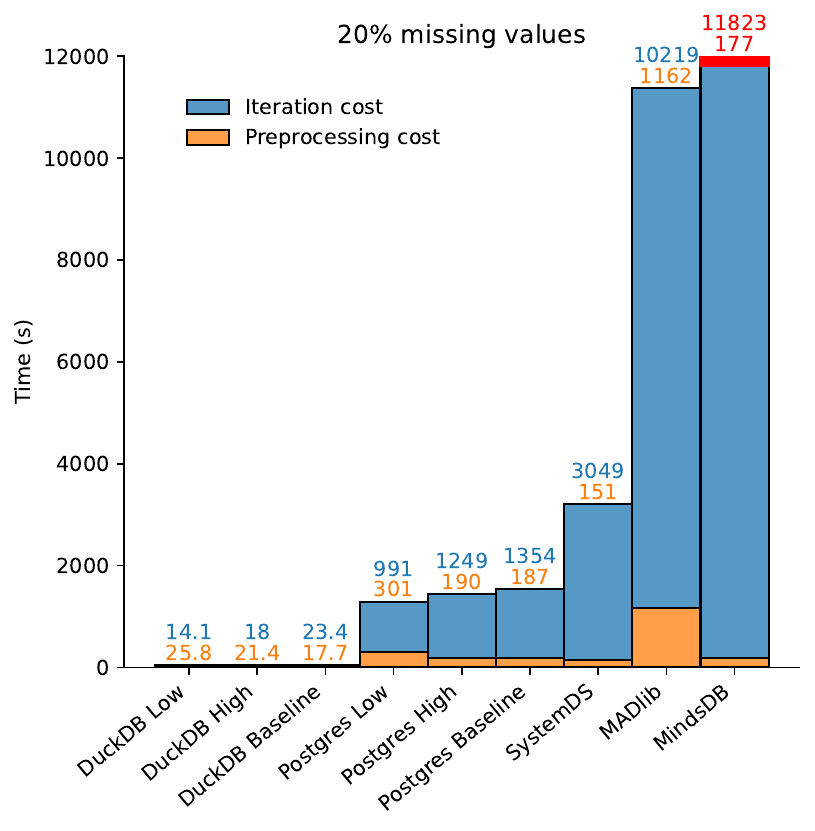}
		\subcaption{Retailer 20\% missing}
		\label{fig:single_table_retailer_0.2}
	\end{minipage}
	
	\begin{minipage}[t]{0.27\textwidth}
		\centering
		\includegraphics[width=\textwidth]{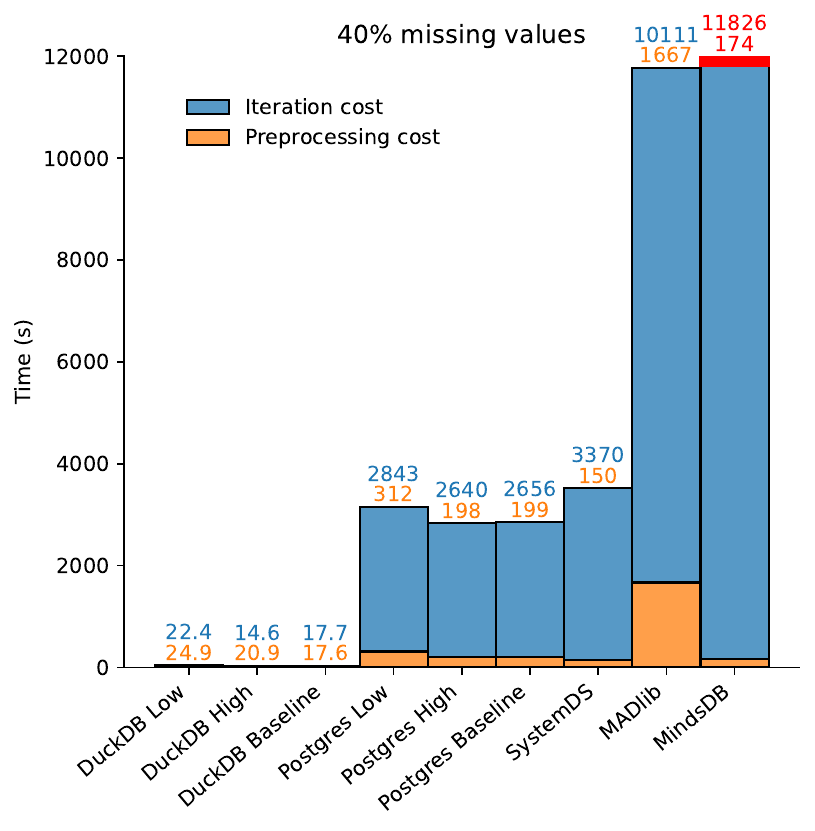}
		\subcaption{Retailer 40\% missing}
		\label{fig:single_table_retailer_0.4}
	\end{minipage}
	\hfill
	\begin{minipage}[t]{0.27\textwidth}
		\centering
		\includegraphics[width=\textwidth]{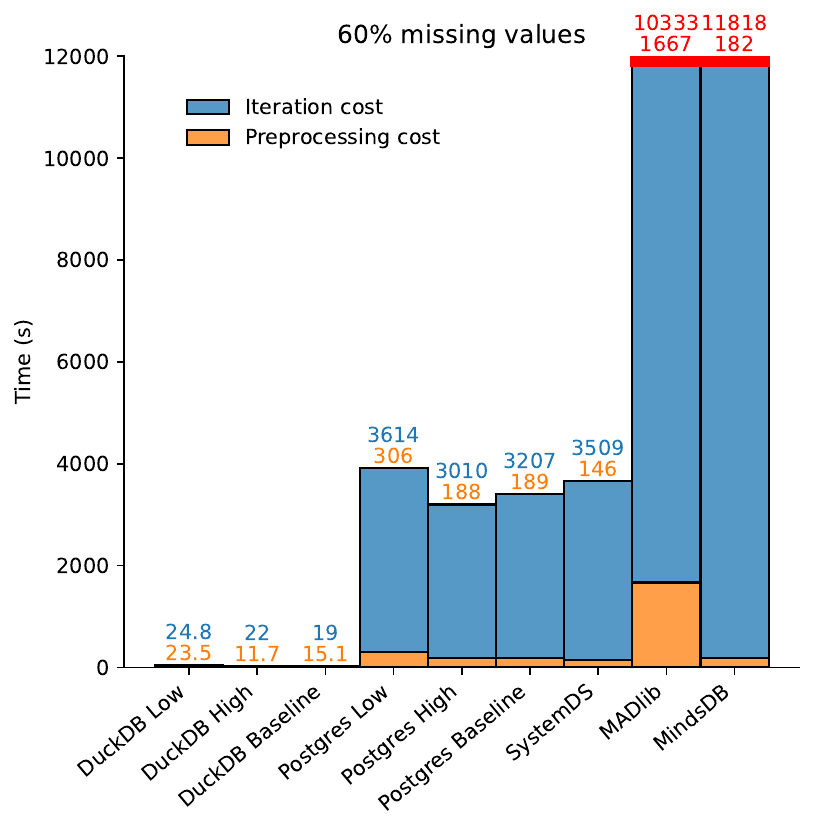}
		\subcaption{Retailer 60\% missing}
		\label{fig:single_table_retailer_0.6}
	\end{minipage}
	\hfill
	\begin{minipage}[t]{0.27\textwidth}
		\centering
		\includegraphics[width=\textwidth]{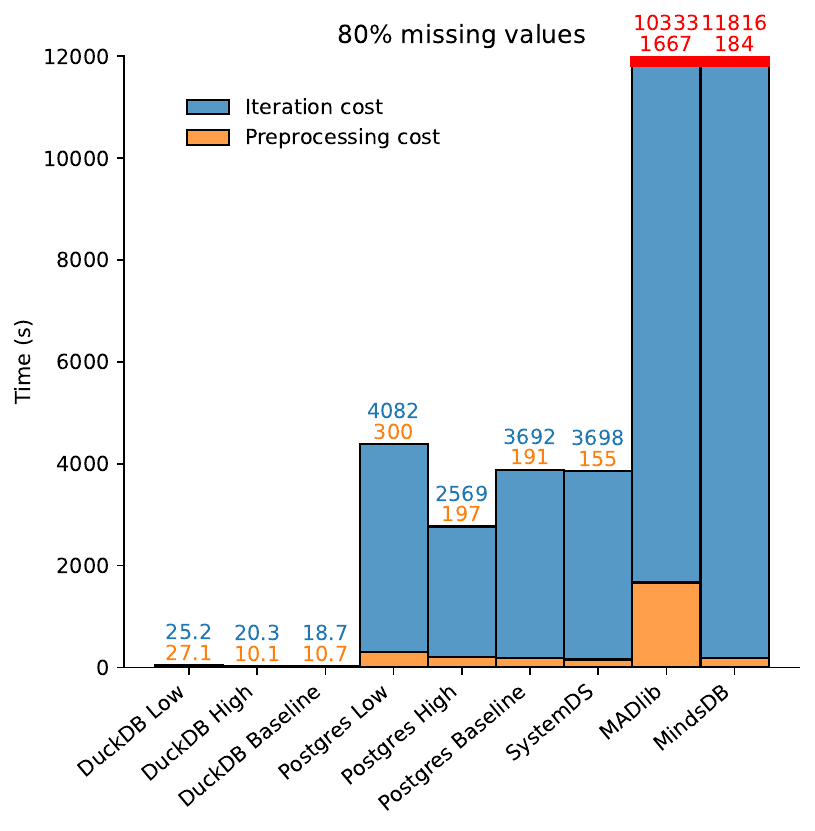}
		\subcaption{Retailer 80\% missing}
		\label{fig:single_table_retailer_0.8}
	\end{minipage}
	\caption{\hl{Single-table imputation using MICE. The time needed to run a single round of MICE over Flight and Retailer with different percentages of missing values, divided into preprocessing time (done once) and iteration time (repeated every round). The baseline, low, and high versions denote our implementations described in Section~\ref{sec:mice_implementation}. The red labels indicate a timeout.}}	
	\label{fig:single_table_imputation}
\end{figure*}

\subsection{Single-Table Imputation}

\hl{We compare our three MICE implementations from Section~\ref{sec:mice_implementation} -- \textsc{Baseline}, \textsc{Low}, and \textsc{High} -- in PostgreSQL and DuckDB against SystemDS, MADlib, and MindsDB, measuring the time required to execute one round of the MICE algorithm, that is, impute values in each incomplete attribute once.} Since every competitor except ours assumes a single table as input, we precompute the join result for the Flight and Retailer datasets. 
We randomly remove different quantities of missing values from 7 columns in each dataset to measure their impact on the runtime. 
\hl{
SystemDS and MADlib implement MICE with linear regression and logistic regression. Both implementations produce predictions similar to ours obtained using stochastic linear regression and LDA; the difference of RMSE on Flight and Retailer is less than 1\% after convergence. MindsDB uses gradient boosting decision trees (LightGBM) for MICE.
}





\hl{
Figure~\ref{fig:single_table_imputation} reports the time required to run a single round of MICE over 7 columns (in blue) and the preprocessing time (in orange) over Flight and Retailer when the percentage of missing values varies between 5\% and 80\%. 
Our baseline implementations in DuckDB and PostgreSQL achieve lower per-iteration costs than all other competitors on both datasets, regardless of the fraction of missing values. 
Compared to the leading competitor, SystemDS, our DuckDB baseline is increasingly faster per iteration as the missing rate increases, ranging from 86x to 176x on Flight and from 106x to 346x on Retailer.  
The performance improvement is due to several reasons. 
All other competitors perform one-hot encoding in the preprocessing stage, increasing the size of the training dataset. SystemDS and MADlib use the direct solve method for linear regression, computing the cofactor matrix and its inverse. 
MindsDB executes expensive training of decision trees for every column. 
Our methods, instead, use gradient descent for regression, computing one compound aggregate in one pass over the dataset, without the need for prior one-hot encoding.
DuckDB further benefits from columnar-vectorized aggregation and inexpensive updates via column swapping. 
The performance advantage of our PostgreSQL baseline over SystemDS decreases with more missing values due to increased update overheads. 
SystemDS and DuckDB exploit the full parallelism of 64 threads, while PostgreSQL does sequential updates and limits its scan parallelism to 8 threads for the two datasets. 

Figure~\ref{fig:single_table_imputation} also shows the effectiveness of our computation sharing techniques (cf. Section~\ref{sec:indb-imputation}). 
The \textsc{Low} implementation, tailored to datasets with low missing rates, reduces the per-iteration baseline cost by 3x in DuckDB and 4.5x in PostgreSQL on the Flight dataset with 5\% of missing values, at the expense of increasing the one-off preprocessing costs; similar holds on Retailer (cf. Figures~\ref{fig:single_table_flight_0.05} and \ref{fig:single_table_retailer_0.05}). The results evidence that the \textsc{Low} implementation pays off on datasets containing up to 20\% missing values. 
The \textsc{High} implementation, although initially designed for datasets with high missing rates, outperforms the baseline in both low and high missing rate scenarios: in the former, this is mainly due to precomputing partial cofactor aggregates over complete records outside the iteration loop, which reduces the iteration cost by 2.7x in DuckDB and 2.4x in PostgreSQL on Flight, and similarly on Retailer; in the latter, this is also due to processing data partitions that are smaller with more missing values, which makes the model training less expensive.
}

\nop{
}


\begin{figure*}[t]
	\begin{minipage}[b]{\textwidth}
		\centering
		\includegraphics[width=0.55\textwidth]{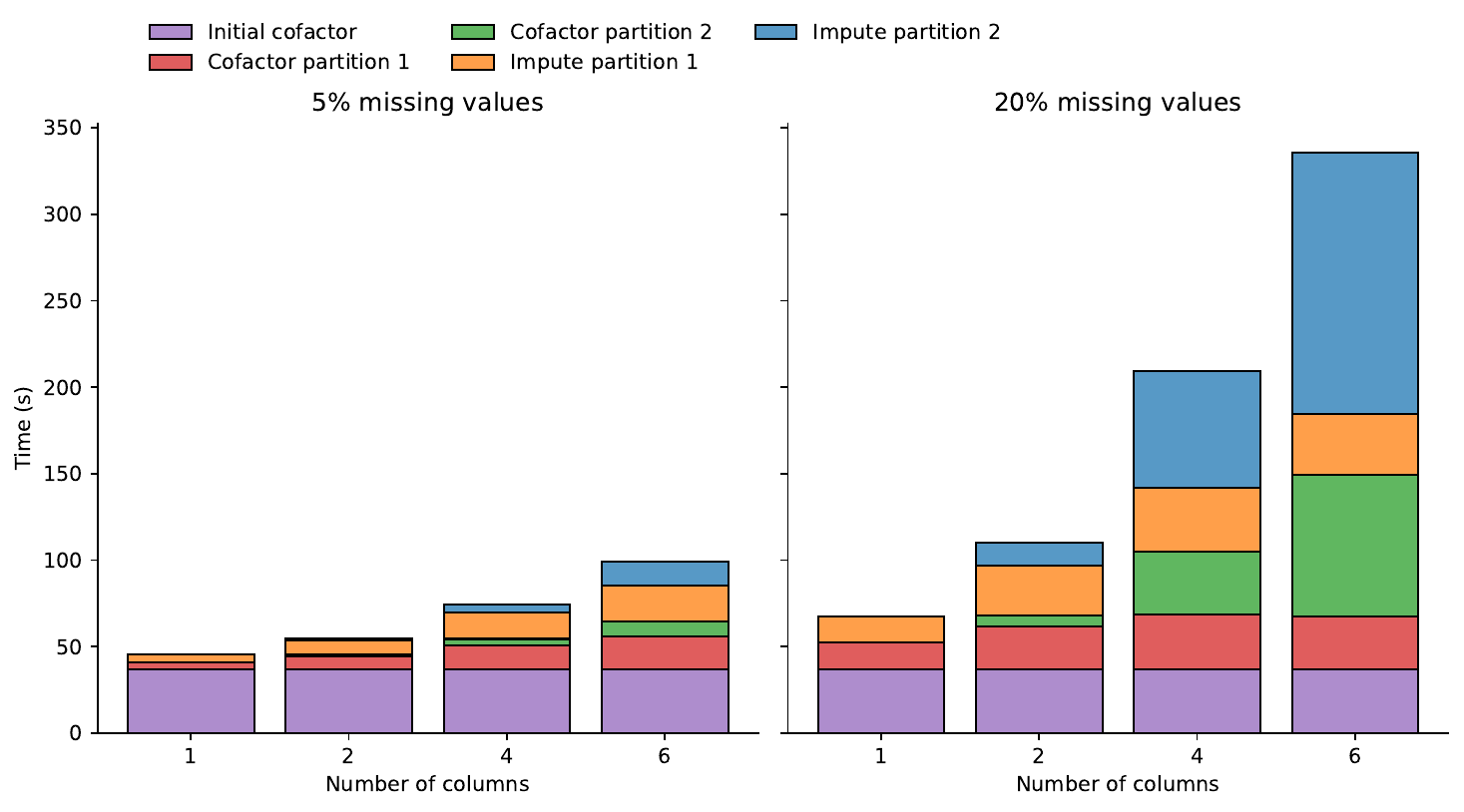}
		\subcaption{PostgreSQL}
		\label{fig:MICE_n_cols_postgres}
	\end{minipage}
	
	\begin{minipage}[b]{\textwidth}
		\centering
		\includegraphics[width=0.55\textwidth]{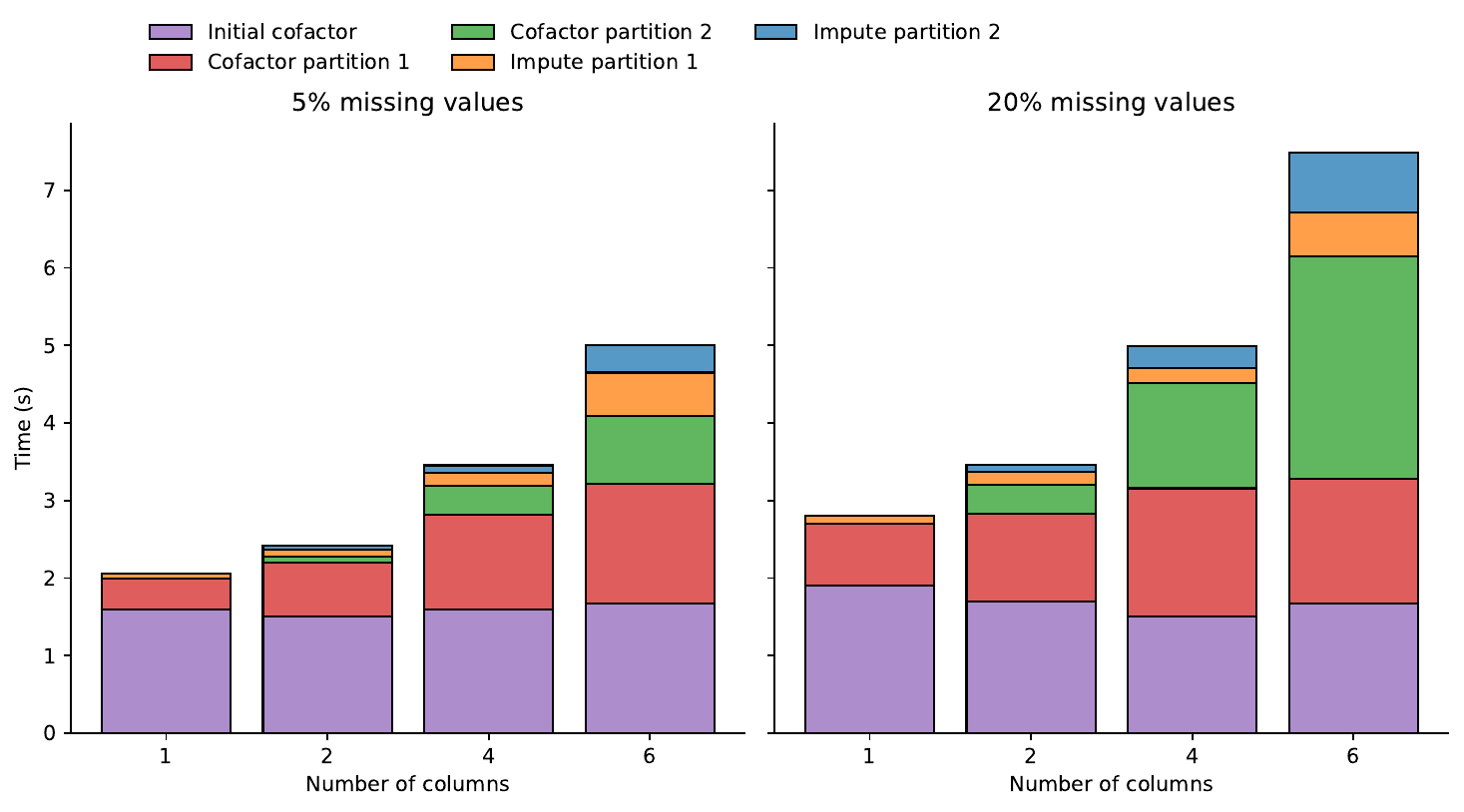}
		\subcaption{DuckDB}
		\label{fig:MICE_n_cols_duckdb}
	\end{minipage}
	\caption{Single-table imputation using the \textsc{Low} implementations with varying numbers of incomplete attributes. The runtime for a single round of MICE on the Flight dataset with randomly generated 5\% and 20\% of missing values in each column.}
	\label{fig:MICE_n_cols}
\end{figure*}

{\bf Varying the Number of Incomplete Attributes.}
We measure the impact of the number of attributes with missing values on the performance of our \hl{\textsc{Low} implementations}. Figure~\ref{fig:MICE_n_cols} reports the breakdown of the runtime for both  PostgreSQL and DuckDB over the Flight dataset as the number of incomplete attributes varies between 1 and 6. The runtime is split into five components: 
the initial time to compute the cofactor aggregates over the entire table,
and, for each of the two affected partitions with missing values,
the time needed to compute the cofactor aggregates in that partition, and
the time needed to update the imputed values in that partition.

For both DBMSs, the figure shows a linear increase in the runtime with respect to the number of attributes with missing values, where the missingness ratio dictates the increase quantity. A higher number of columns with missing values also increases the size of the partition containing rows with at least two missing values, prolonging also the cofactor computation and imputation time.

PostgreSQL and DuckDB have different bottlenecks: in the former, updating imputed values is the slowest phase, while in the latter, it is the computation of cofactor aggregates. This is due to their different architectures. PostgreSQL uses a row-based engine with MVCC, which despite our optimizations, still causes significant overheads during the execution of update queries. DuckDB, instead, adopts a column-oriented approach, which we exploit to implement lightweight column swapping to instantiate new imputed values.

\begin{figure*}[t]
	\begin{minipage}[b]{\textwidth}
		\centering
		\includegraphics[width=0.9\textwidth]{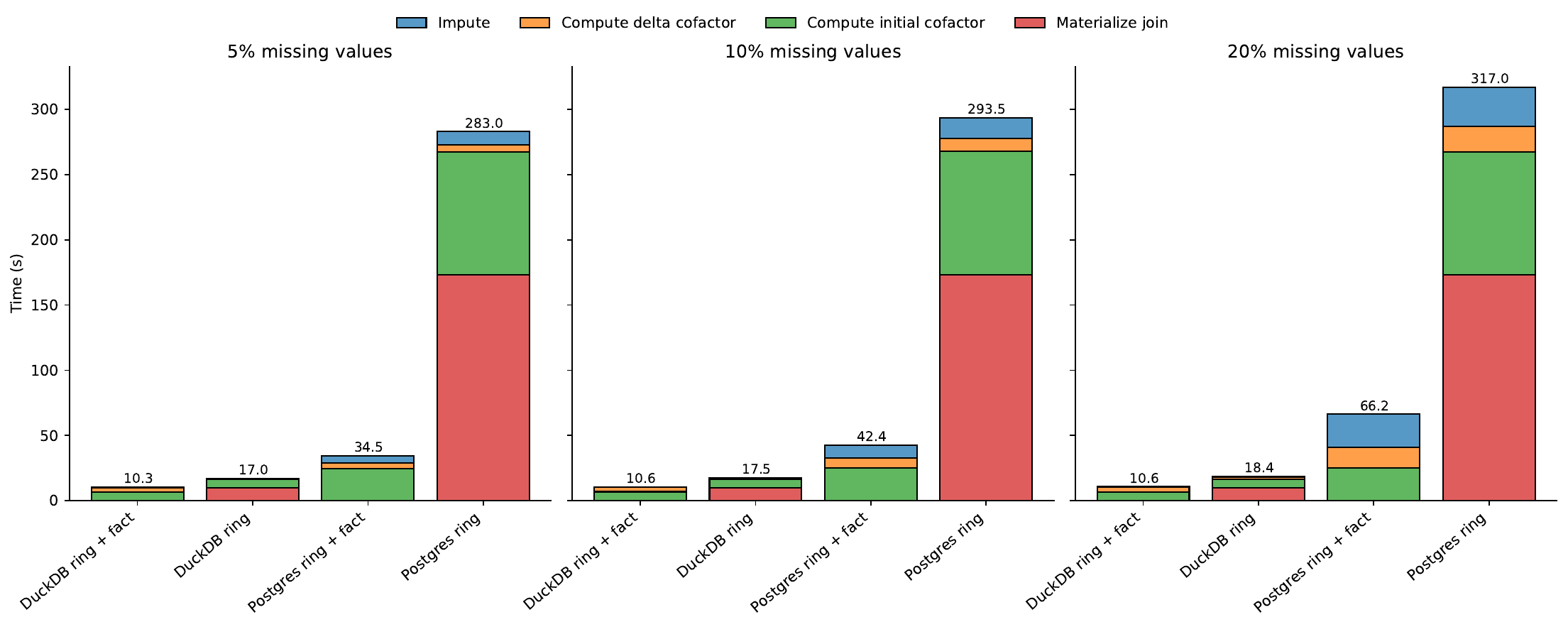}
		\subcaption{Retailer dataset}
		\label{fig:retailer_factorized}
	\end{minipage}
	
	\begin{minipage}[b]{\textwidth}
		\centering
		\includegraphics[width=0.9\textwidth]{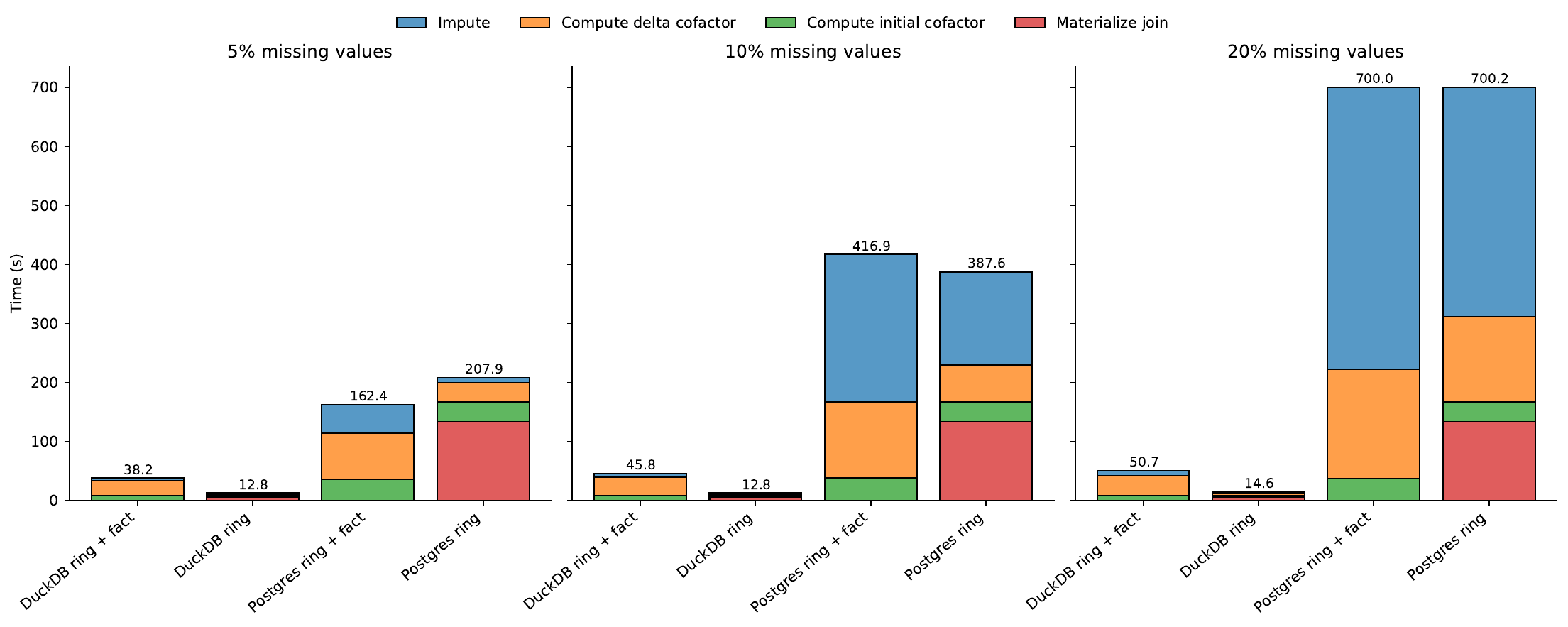}
		\subcaption{Flight dataset}
		\label{fig:flight_factorized}
	\end{minipage}
	\caption{Imputation over normalized data using the \textsc{Low} implementations with materialization of the join result and with factorized evaluation that avoids the join materialization. 
	The runtime for a single round of MICE with DuckDB and PostgreSQL over Retailer and Flight with different percentages of missing values, randomly generated in the fact table of each dataset.}
	\label{fig:exp_factorized}
\end{figure*}

\subsection{Imputation over Normalized Data}

We now consider the case when the dataset is normalized, and the imputation takes into account the attributes from all tables. 
We compare the performance of our \hl{{\textsc{Low}} implementations} running over the materialized join result and its factorized version that pushes the aggregate computation past the joins. We use the Flight and Retailer datasets as they consist of multiple tables, and randomly generate missing values in the fact table only so that in both cases the algorithm generates the same imputations; otherwise, joining  incomplete tables can create copies of the same missing value, which might be imputed differently by the two approaches. 


Figure~\ref{fig:exp_factorized} shows the performance of the two approaches over a single MICE round for Retailer and Flight, consisting of 1 and 7 attributes with missing values, respectively. As previously discussed, in the Flight dataset, joining its tables does not introduce many redundant values, thus imputing values over a normalized database is slower than using the joined table.
%
On the Retailer dataset, instead, we get the opposite result. As the joined result is more redundant, imputing values directly over normalized relations in PostgreSQL is from 4.8 to 8.2x faster than using the joined result, mainly because of the long time to materialize the joined table and compute the cofactor aggregates over redundant values. The imputation time is also faster: PostgreSQL writes a full row when a new value is imputed, and updating a table with just 4 attributes is faster than updating 25 attributes of the joined table. \hl{DuckDB also imputes faster over the normalized dataset than over the joined dataset, albeit with a smaller improvement of 1.7x than PostgreSQL, mainly due to faster in-memory join materialization and aggregation. }

\begin{figure}[t]
	\begin{minipage}[]{0.33\textwidth}
		\centering
		\includegraphics[width=\textwidth]{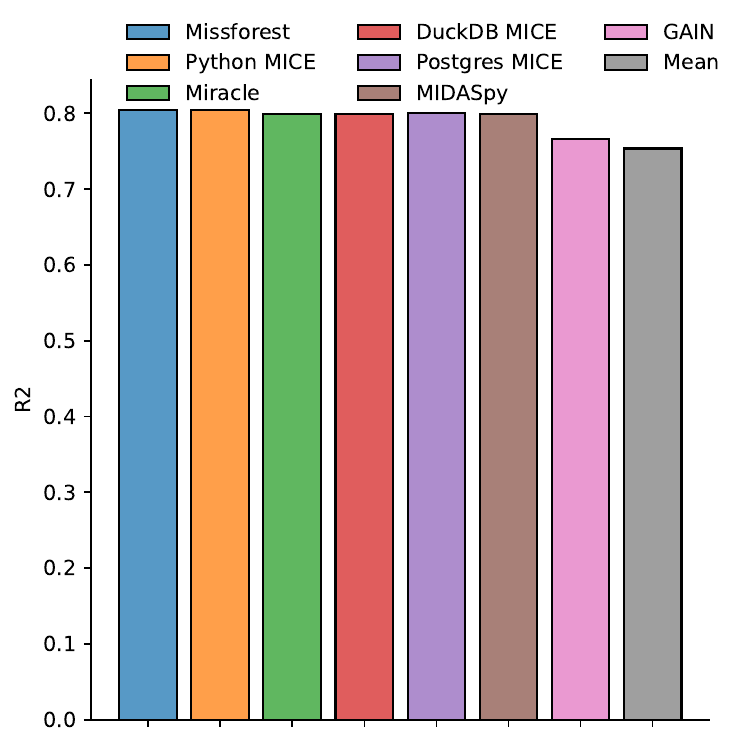}
		\subcaption{R2}
		\label{fig:air_quality_r2}
	\end{minipage}
	\hfill
	\begin{minipage}[]{0.33\textwidth}
		\centering		
		\includegraphics[width=\textwidth]{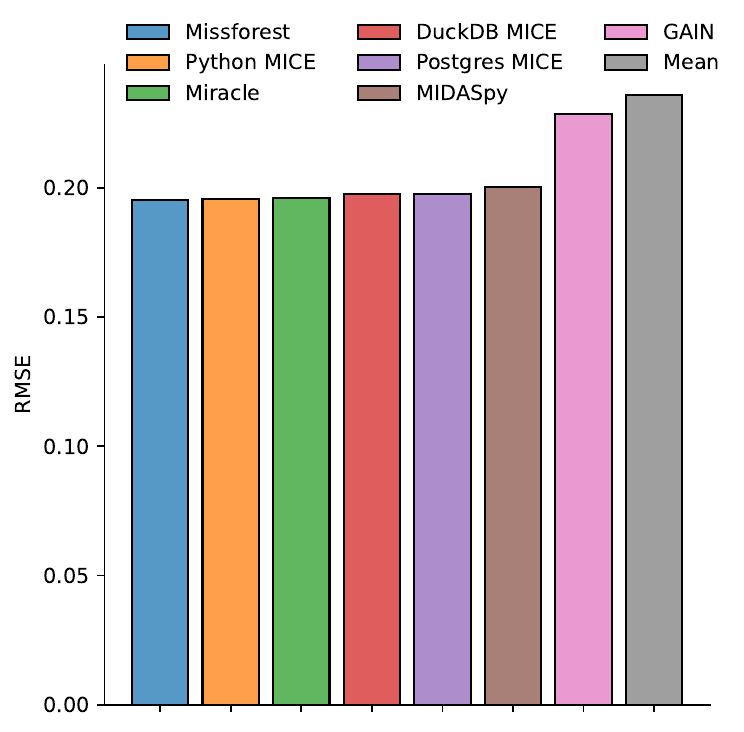}
		\subcaption{RMSE}
		\label{fig:air_quality_mse}
	\end{minipage}
	\hfill
	\begin{minipage}[]{0.30\textwidth}
		\centering
		\includegraphics[width=\textwidth]{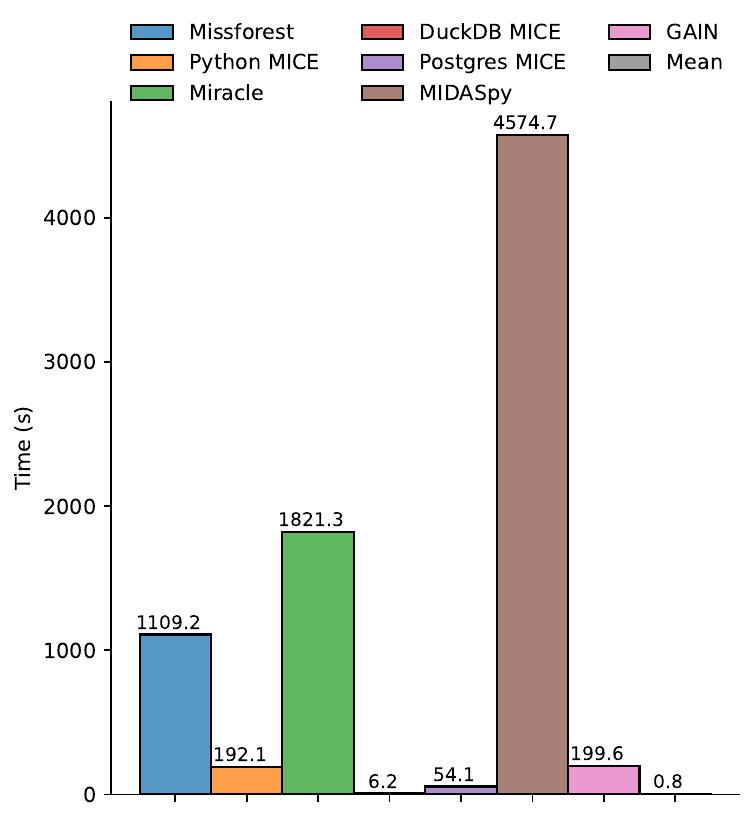}
		\subcaption{Imputation time}
		\label{fig:air_quality_time_python}
	\end{minipage}
	
	\caption{Imputation quality on the Air Quality dataset. The quality of a regression model that predicts the air quality index given attributes with missing values, measured by R2 (higher is better) and RMSE (lower is better). The time needed to impute the dataset using different imputation approaches.}	
	\label{fig:air_quality_imp_quality}
\end{figure}

\subsection{Imputation Quality}

\hl{
We compare different imputation methods in terms of imputation quality, measured as the performance of a linear regression model trained over imputed data.
We consider the following methods:
MICE DuckDB and MICE PostgreSQL, our MICE implementations (\textsc{Low} versions);
MICE Python, the MICE implementation from scikit-learn using linear and logistic regression;
MissForest~\cite{Stekhoven:2012:MissForest}, which uses random forests;
GAIN~\cite{Yoon:2018:GAIN}, which uses generative adversarial networks;
MIRACLE~\cite{Kyono:2021:Miracle}, which uses causally-aware refinements,
MIDASpy~\cite{Ranjit:2022:MIDAS}, which uses autoencoders;
and mean/mode imputation. 
We use the Python implementation of the competitors from HyperImpute~\cite{Jarrett:2022:HyperImpute}. 
We use 5 iterations for the MICE methods.

Figure~\ref{fig:air_quality_imp_quality} shows the imputation quality over the Taiwan's Air Quality dataset, where the task is to predict the air quality index given its pollutants. 
Although the dataset contains only 6\% of missing values,
mean imputation performs significantly worse than the other methods, reducing the R2 score of the regression model by 0.05 and increasing its root mean squared error (RMSE) by 0.05. 
MICE outperforms GAIN and achieves similar performance to MissForest, MIDASpy, and MIRACLE.
Figure~\ref{fig:air_quality_time_python} reports the imputation time of each method.
MICE DuckDB is faster than 
GAIN by 32x,
MICE Python by 38x, 
MIDASpy by 778x, 
and only 5.4 seconds slower than mean imputation in Python.

\begin{figure*}
	\begin{minipage}[]{\linewidth}
		\centering\footnotesize
		\resizebox{\textwidth}{!}{
		\begin{tabular}{@{}l@{\hspace{2mm}}lc@{\hspace{1.2mm}}c@{\hspace{1.2mm}}c@{\hspace{1.2mm}}c@{\hspace{1.2mm}}c@{\hspace{1.2mm}}cc@{}r@{}r@{~}}
			\toprule
			\multirow{3}{*}{\bf Pattern} & \multirow{3}{*}{\bf Method} & \multicolumn{6}{@{}c@{}}{\bf Missing Rate} & &
			\multicolumn{2}{c}{\bf Imp. Time}\\
			\cmidrule{3-8}\cmidrule{10-11}
			& & 0.05 & 0.1 & 0.2 & 0.4 & 0.6 & 0.8 & & secs & norm \\
			\midrule
			\multirow{7}{*}{\bf MCAR}
				& MICE DuckDB & 0.180 $\pm$ 0.009& 0.181 $\pm$ 0.002 & 0.161 $\pm$ 0.005 & 0.175 $\pm$ 0.016 & 0.173 $\pm$ 0.012 & 0.175 $\pm$ 0.021 & & 5.4 & 1.0 \\
				& MICE Python & 0.180 $\pm$ 0.005 & 0.183 $\pm$ 0.002& {\bf 0.158} $\pm$ 0.003 & 0.174 $\pm$ 0.010 & 0.172 $\pm$ 0.012 & 0.173 $\pm$ 0.018 & & 189.6 & 34.9 \\
				& Mean & 0.286 $\pm$ 0.009& 0.359 $\pm$ 0.009 & 0.473 $\pm$ 0.008 & 0.642 $\pm$ 0.025 & 0.774 $\pm$ 0.012 & 0.889 $\pm$ 0.012 & & 0.4 & 0.1 \\
				& MissForest & 0.235 $\pm$  0.004 & 0.275 $\pm$  0.011 & 0.333 $\pm$  0.005& 0.431 $\pm$  0.021 & 0.476 $\pm$  0.005 & 0.461 $\pm$  0.070 & & 407.5 & 75.0 \\
				& GAIN & 0.286 $\pm$ 0.038 & 0.350 $\pm$ 0.001 & 0.460 $\pm$ 0.015 & 0.633 $\pm$ 0.019 & 0.756 $\pm$ 0.020 & 0.864 $\pm$  0.129 & & 82.0 & 15.1 \\
				& MIRACLE & {\bf 0.179} $\pm$ 0.003 & {\bf 0.180}  $\pm$ 0.001 & 0.175 $\pm$ 0.003 &  {\bf 0.169} $\pm$ 0.002 & {\bf 0.166} $\pm$ 0.001 & {\bf 0.168} $\pm$ 0.002 & & 788.4 & 145.1 \\
			\midrule
			\multirow{7}{*}{\bf MAR}
				 & MICE DuckDB & 0.176 $\pm$ 0.021 & 0.175 $\pm$ 0.043 &  0.268 $\pm$ 0.013 & {\bf 0.336} $\pm$ 0.078 & {\bf 0.331} $\pm$ 0.009 & 0.344 $\pm$ 0.068 & & 5.5 & 1.0 \\
				 & MICE Python & {\bf 0.174} $\pm$ 0.016 & {\bf 0.172} $\pm$ 0.039 & {\bf 0.266} $\pm$ 0.012 & 0.337 $\pm$ 0.071 & 0.339 $\pm$ 0.012 & {\bf 0.342} $\pm$ 0.038 & & 186.6 & 34.2 \\
				 & Mean & 0.487 $\pm$ 0.035 & 0.597 $\pm$  0.020 & 0.723 $\pm$ 0.001 & 0.845 $\pm$ 0.005 & 0.913 $\pm$ 0.002 & 0.958 $\pm$ 0.004 & & 0.4 & 0.1 \\
				 & MissForest & 0.367 $\pm$ 0.003 & 0.431 $\pm$ 0.013 & 0.510 $\pm$ 0.001 & 0.557 $\pm$ 0.095 & 0.499 $\pm$  0.006 & 0.486 $\pm$ 0.081 & & 415.5 & 76.1 \\
				 & GAIN & 0.384 $\pm$ 0.137 & 0.485 $\pm$ 0.092 & 0.674 $\pm$ 0.129 & 0.868 $\pm$ 0.133 & 0.917 $\pm$ 0.179 & 0.923 $\pm$ 0.001 & & 81.0 & 14.8 \\
				 & MIRACLE & 0.191 $\pm$ 0.025 & 0.217 $\pm$ 0.036 & 0.257 $\pm$ 0.007 & 0.397 $\pm$ 0.077 & 0.538 $\pm$ 0.044 & 0.632 $\pm$ 0.536 & & 785.6 & 143.9 \\
			\midrule
			\multirow{7}{*}{\bf MNAR}
				 & MICE DuckDB & 0.194 $\pm$ 0.011 & 0.193 $\pm$ 0.018 & 0.199 $\pm$ 0.037 & 0.280 $\pm$ 0.023& 0.203 $\pm$ 0.041& 0.231 $\pm$ 0.068& & 5.5 & 1.0 \\
				 & MICE Python & 0.192 $\pm$ 0.008 & 0.195 $\pm$ 0.021 & 0.184 $\pm$ 0.029 & 0.279 $\pm$ 0.021 & 0.210 $\pm$ 0.043 & 0.229 $\pm$ 0.053 & & 186.2 & 33.9\\
				 & Mean & 0.289 $\pm$ 0.012 & 0.365 $\pm$ 0.034 & 0.475 $\pm$ 0.004 & 0.651 $\pm$ 0.065 & 0.779 $\pm$ 0.039 & 0.893	$\pm$ 0.039 & & 0.4 & 0.1 \\
				 & MissForest & 0.239 $\pm$ 0.014 & 0.280 $\pm$ 0.010 & 0.347 $\pm$ 0.073 & 0.436 $\pm$ 0.017 & 0.487 $\pm$ 0.045 & 0.490 $\pm$ 0.086 & & 553.6 & 100.8 \\
				 & GAIN & 0.284 $\pm$ 0.039 & 0.355 $\pm$ 0.089 & 0.463 $\pm$ 0.048 & 0.633 $\pm$ 0.058 & 0.759 $\pm$ 0.110 & 0.898 $\pm$ 0.110 & & 79.4 & 14.5 \\
				 & MIRACLE & {\bf 0.186} $\pm$ 0.007  & {\bf 0.191} $\pm$ 0.037& {\bf 0.182} $\pm$ 0.009 & {\bf 0.185} $\pm$ 0.099 & {\bf 0.200} $\pm$ 0.099 & {\bf 0.223} $\pm$ 0.102& & 795.4 & 144.8 \\
			\bottomrule
		\end{tabular}}
		\vspace{6pt}
		\subcaption{Flight}
		\vspace{12pt}
		\label{fig:flight_patterns_mse}
	\end{minipage}

	\begin{minipage}[]{\linewidth}
		\centering\footnotesize
		\resizebox{\textwidth}{!}{
		\begin{tabular}{@{}l@{\hspace{2mm}}lc@{\hspace{1.2mm}}c@{\hspace{1.2mm}}c@{\hspace{1.2mm}}c@{\hspace{1.2mm}}c@{\hspace{1.2mm}}cc@{}r@{}r@{~}}
			\toprule
			\multirow{3}{*}{\bf Pattern} & \multirow{3}{*}{\bf Method} & \multicolumn{6}{@{}c@{}}{\bf Missing Rate} & &
			\multicolumn{2}{c}{\bf Imp. Time}\\
			\cmidrule{3-8}\cmidrule{10-11}
			& & 0.05 & 0.1 & 0.2 & 0.4 & 0.6 & 0.8 & & secs & norm \\
			\midrule
			\multirow{7}{*}{\bf MCAR}
				& MICE DuckDB & {\bf 0.228}  $\pm$ 0.013 & 0.241 $\pm$ 0.024 & 0.249 $\pm$ 0.029 & 0.254 $\pm$ 0.009 & 0.283 $\pm$ 0.036 & 0.296 $\pm$ 0.013 & & 5.7 & 1.0 \\
				& MICE Python & 0.233 $\pm$ 0.009 & {\bf 0.237} $\pm$ 0.016 & {\bf 0.248} $\pm$ 0.026 & {\bf 0.252} $\pm$ 0.004 & {\bf 0.282} $\pm$ 0.038 & {\bf 0.294} $\pm$ 0.015 & & 128.3 & 22.6 \\
				& Mean & 0.316 $\pm$ 0.015 & 0.382 $\pm$ 0.045 & 0.491 $\pm$ 0.015 & 0.645 $\pm$ 0.010 & 0.755 $\pm$ 0.022 & 0.841 $\pm$ 0.011 & & 0.8 &  0.1 \\
				& MissForest & 0.267 $\pm$ 0.011 & 0.304 $\pm$ 0.002 & 0.361 $\pm$ 0.028 & 0.446 $\pm$ 0.001 & 0.519  $\pm$ 0.046 & 0.558 $\pm$ 0.010 & & 212.7 & 37.5 \\
				& GAIN & 0.294 $\pm$ 0.003& 0.357 $\pm$ 0.149 & 0.449 $\pm$ 0.103 & 0.571 $\pm$ 0.128 & 0.673 $\pm$ 0.253 & 0.735 $\pm$ 0.010 & & 81.0 & 14.3 \\
				& MIRACLE & 0.268 $\pm$ 0.044 & 0.320 $\pm$ 0.102 & 0.378 $\pm$ 0.113 & 0.611 $\pm$ 0.133 & 0.749 $\pm$  0.215 & 0.791 $\pm$  0.271 & & 995.1 & 174.5\\
			\midrule
			\multirow{7}{*}{\bf MAR}
			& MICE DuckDB & {\bf 0.226} $\pm$ 0.006 & 0.242 $\pm$ 0.009& {\bf 0.245} $\pm$ 0.023 & 0.261 $\pm$ 0.019 & 0.277 $\pm$ 0.016 & {\bf 0.296} $\pm$ 0.028 & & 5.7 & 1.0 \\
			& MICE Python & 0.234 $\pm$ 0.005 & {\bf 0.239} $\pm$ 0.008 & 0.247 $\pm$ 0.019 & {\bf 0.260} $\pm$ 0.019 & {\bf 0.275} $\pm$ 0.013 & 0.299 $\pm$ 0.029 & & 125.7 & 22.1 \\
			& Mean & 0.316 $\pm$ 0.022 & 0.384 $\pm$ 0.057 & 0.489 $\pm$ 0.026 & 0.638 $\pm$ 0.007 & 0.740 $\pm$ 0.028 & 0.841 $\pm$ 0.027 & & 0.8 & 0.1 \\
				& MissForest & 0.268 $\pm$ 0.008 & 0.303 $\pm$ 0.006 & 0.360 $\pm$  0.026 & 0.449 $\pm$ 0.029 & 0.520 $\pm$ 0.009 & 0.551 $\pm$ 0.031 & & 227.1 & 39.9 \\
				& GAIN & 0.294 $\pm$ 0.019 & 0.352 $\pm$ 0.073& 0.459 $\pm$ 0.025& 0.595 $\pm$ 0.171 & 0.670 $\pm$ 0.054& 0.744 $\pm$ 0.031 & & 78.4 & 13.8 \\
				& MIRACLE & 0.251 $\pm$ 0.075 & 0.378 $\pm$ 0.167 & 0.342 $\pm$ 0.021 & 0.595 $\pm$ 0.219& 0.783 $\pm$ 0.013 & 0.823 $\pm$ 0.219 & & 984.1 & 173.8\\
			\midrule
			\multirow{7}{*}{\bf MNAR}
				& MICE DuckDB & {\bf 0.227} $\pm$ 0.061 & 0.239 $\pm$ 0.073 & {\bf 0.245} $\pm$ 0.061& 0.258 $\pm$ 0.069 & {\bf 0.279} $\pm$ 0.018 & 0.296 $\pm$ 0.037 & & 5.8 & 1.0 \\
				& MICE Python & 0.232 $\pm$ 0.056 & {\bf 0.238} $\pm$ 0.069 & 0.249 $\pm$ 0.065 & {\bf 0.256} $\pm$ 0.064 & {\bf 0.279} $\pm$ 0.015 & {\bf 0.294} $\pm$ 0.034 & & 127.0 & 22.0 \\
				& Mean & 0.309 $\pm$ 0.149 & 0.365 $\pm$ 0.176 & 0.468 $\pm$ 0.152 & 0.615 $\pm$ 0.184 & 0.698 $\pm$ 0.136 & 0.824 $\pm$ 0.191 & & 0.8 & 0.1 \\
				& MissForest & 0.263 $\pm$ 0.081 & 0.296 $\pm$ 0.133 & 0.350 $\pm$ 0.165 & 0.429 $\pm$ 0.167 & 0.487 $\pm$ 0.021 & 0.539 $\pm$ 0.166 & & 255.4 & 44.3 \\
				& GAIN & 0.283 $\pm$ 0.053 & 0.337 $\pm$ 0.056 & 0.423 $\pm$ 0.229 & 0.552 $\pm$ 0.280 & 0.656 $\pm$ 0.041 & 0.732 $\pm$ 0.018 & & 78.7 & 13.6 \\
				& MIRACLE & 0.228 $\pm$ 0.075 & 0.256 $\pm$ 0.039 & 0.441 $\pm$ 0.104 & 0.601 $\pm$ 0.161 & 0.670 $\pm$ 0.159 & 0.850 $\pm$ 0.156 & & 997.1 & 172.3 \\
			\bottomrule
		\end{tabular}}
		\vspace{6pt}
		\subcaption{Retailer}
		\label{fig:retailer_patterns_mse}
	\end{minipage}

	\caption{\hl{Imputation quality and runtime for different missing rates and patterns on the restricted Flight and Retailer datasets. The quality is measured by the RMSE of a regression model predicting flight duration for Flight and inventory stock for Retailer from the imputed data. The imputation times (absolute and normalized to MICE DuckDB) are for the missing rate of 20\%. }
	}
	\label{fig:missing_patterns_quality}
\end{figure*}

{\bf Missing Patterns.}
We analyze imputation quality under three widely-used mechanisms for injecting missing values:
missing completely at random (MCAR), missing at random (MAR), and missing not at random (MNAR)~\cite{Little:2019:Book}. 
To allow all competitors to finish within a 30-minute timeout, 
we restrict the Flight dataset to year 2015 (5M rows) and the Retailer dataset to two dates with ids 501 and 508 (1M rows).
For the former, the task is to predict flight duration given predictors (e.g., distance between airports); 
for the latter, the task is to predict inventory stock given predictors (e.g., population in the area).

We use the generator from HyperImpute~\cite{Jarrett:2022:HyperImpute} to inject missing values in 7 attributes in each dataset. With MCAR, we generate missing values with a uniform distribution. With MAR, the fraction of missing values depends on flight duration in Flight and inventory stock in Retailer. With MNAR, we generate missing values taking all 7 incomplete attributes as input for each dataset.

Figure~\ref{fig:missing_patterns_quality} reports the performance of these imputation methods on the two restricted datasets. 
On the Flight dataset, the MICE methods and MIRACLE outperform the other methods in terms of quality, achieving the lowest RMSEs under all three patterns and all missing rates. But in terms of imputation time, MICE DuckDB is 145x faster than MIRACLE and 35x faster than MICE Python.
The deep learning approaches, GAIN and MIDASpy, generate increasingly worse imputations as the fraction of missing values increases.
Mean imputation performs the worst in terms of imputation quality in all scenarios. 
We observe similar trends on the Retailer dataset in terms of MICE's runtime performance and imputation quality.
}

\section{Related Work}
\textbf{Data Imputation.} 
Simple techniques for handling missing data, such as removing tuples with missing values, mean/mode imputation, indicator imputation, and Last Observation Carried Forward~\cite{Lachin:2016:LOCF}, are fast but only work under restrictive assumptions, produce biased analytics, and might distort the value distribution.

\hl{
Model-based imputation aims to address these issues. 
This includes approaches that learn the joint distribution of data based on matrix completion~\cite{Mazumder:2010:Spectral}, generative adversarial networks~\cite{Yoon:2018:GAIN, Yoon:2020:GAMIN}, diffusion models~\cite{Zheng:2022:Diffusion}, autoencoders~\cite{Peis:2022:Missing, Mattei:2019:MIWAE}, and the EM algorithm~\cite{Little:2019:Book}. 
Discriminative approaches to data imputation include deep denoising autoencoders~\cite{Ranjit:2022:MIDAS}, graph neural networks~\cite{You:2020:Missing}, and conditional modeling strategies such as MICE~\cite{vanBuuren:2011:miceR} and MissForest~\cite{Stekhoven:2012:MissForest}.
Specific imputation methods are designed for particular domains, such as time-series~\cite{Parikshit:2021, Xiaoou:2019:Cleanits, Mourad:2020:ORBITS, Khayati:2020:VLDB} and non-numerical data~\cite{Biessmann:2018:DeepLearning}. 
Imputation methods can also be based on multiple models: 
HoloClean~\cite{Rekatsinas:2017:HoloClean} does broader data cleaning using statistical analysis, integrity constraints, and external data, while MIRACLE~\cite{Kyono:2021:Miracle} learns both imputation function and missingness graph.
}

\hl{While these methods offer high imputation quality, they also require external tools, can only impute values in a single table, and have a long imputation time, which hinders their applicability on large datasets: most of them consider datasets with at most few hundreds of thousands records, while their imputation times range from 1.7K to 96K seconds on a 1M dataset~\cite{Miao:2021:ImputeInfluenceFn}. 
Some systems try to avoid lengthy imputation. SampleClean~\cite{Krishnan:2015} estimates the result of a query over dirty data by cleaning only a sample of the dataset. ImputeDB~\cite{Cambronero:2017} incorporates the imputation process into the query optimizer, imputing only the relevant data; this can generate biased models.
EDIT~\cite{Miao:2021:ImputeInfluenceFn} addresses the high cost of training deep learning models by reducing the training dataset, estimating the importance of each sample. Such importance estimates, however, are often erroneous for deep networks~\cite{Basu:2021:InfluenceFn}}.

Compared to simple data imputation techniques, our approach offers a similar runtime while providing significantly better imputation quality. With respect to model-based imputation tools, our approach avoids the problem of moving data between different systems, imputes data directly over multiple tables, and significantly improves the imputation time. Compared to hybrid systems such as \hl{SampleClean, ImputeDB and EDIT, our approach does not reduce the dataset size to improve performance and does not require preprocessing such as joining relations or one-hot encoding.
}

\textbf{Systems for Treating Missing Values.}
\hl{There are systems that try to adapt tasks to work over incomplete databases, such as entity resolution~\cite{Weilong:2021:EntityRes}, discovery of functional dependencies~\cite{Berti:2018:DiscoveryFD}, or machine learning training~\cite{Picado:2020:LearningDirty, Karla:2020:NNC, Tongyu:2021:Adaptive}.
Other systems try to repair missing values using different approaches, such as human supervision~\cite{Mashaal:2020:CoClean, Mahdavi:2020:Baran}, functional dependencies~\cite{Rezig:2021:Horizon}, ensemble of tools~\cite{Ziawasch:2016:DetectErr}, by generating interpretable data repair rules~\cite{Peng:2022:GAN}, or synthesizing completely missing tuples over database schema~\cite{Hilprecht:2021:Restore}. 
Missing values might also not be clearly recognizable, prompting systems for the discovery of hidden missing values~\cite{Abdulhakim:2018:FAHES} and the evaluation of the quality of incomplete databases~\cite{Liang:2020, Schelter:2018:Verification}.
}

\textbf{In-Database Learning.}
This line of work aims to enable machine learning tasks within a database system, delivering faster training and prediction times, lower data movement and storage costs, and better data security. 
MADLib~\cite{Hellerstein:2012:MADlib} and Bismarck~\cite{Feng:2012:Bismarck} implement ML tasks inside user-defined functions, thus removing the need for data movementm, but without exploiting the structure of the data to improve learning performance. 
%
LMFAO~\cite{Schleich:2019:LMFAO} and AC/DC~\cite{Khamis:2018:ACDC} support learning various ML models over normalized tables by sharing computation, Orion~\cite{Kumar:2015} supports generalized linear models, and Morpheus~\cite{Chen:2017:Morpheus} supports linear algebra operations over a normalized database. F-IVM~\cite{Nikolic:2018:SIGMOD,Nikolic:2020:FIVM} introduces the cofactor ring to support incremental computation of ML models.

Compared to this prior work, we make several novel contributions: 
we present a new method for in-database training of linear discriminant analysis models, 
we adapt the widely-used MICE algorithm to exploit computation sharing, and 
instead of developing specialized tools, we implement our approach in two popular database systems, offering high data imputation quality at low cost.

\section{Conclusion}
This paper introduces a novel and efficient method for generating high-quality data imputation.
We show how to enhance the widely-used MICE algorithm to exploit computation sharing and scalable execution,
tailoring it specifically for the implementation within a database system.
We implement our imputation methods in DuckDB and PostgreSQL  
and demonstrate their superiority when compared to other methods,
resulting in faster imputation time without compromising the quality of imputed data.

Looking ahead, we plan to enrich our open-source library with new methods for in-database learning and data cleaning, offering data practitioners a powerful tool to effectively address the challenges presented by missing data. 
Our library already supports other ML algorithms not elaborated in this paper, such as Na\"{i}ve Bayes and Quadratic Discriminant Analysis classifiers. 
We believe it is possible to further empower in-database data cleaning tasks with learning algorithms such as decision trees, k-means, random forests, and gradient boosting~\cite{Huang:2023:JoinBoost, Schleich:2020, Chen:2017:Morpheus}, including their integration into MICE, albeit the opportunities for computation sharing across models and iterations might not always exist.  
In that context, we want to pursue support for shared and incremental learning of models in iterative database tasks, like in this work. 
Going further, deep learning and deep generative models pose a significant challenge at the moment for a fully in-database implementation due to their non-linearity, thus efficiently training such high-capacity models over relational databases stands as an open problem. 

\bibliographystyle{abbrv}
\bibliography{references}


\end{document}